\documentclass[12pt,aps,prb,preprint,showpacs,,superscriptaddress]{revtex4}
\usepackage{graphicx,amssymb,bm,makeidx}
\usepackage{amsfonts}
\usepackage{amsmath,amssymb}
\usepackage{grffile}
\usepackage{xcolor}
\usepackage[normalem]{ulem}
\usepackage{soul}
\usepackage{ulem}
\usepackage{hyperref} 
\hypersetup{colorlinks=true,
            linkcolor=blue,
            citecolor=blue,
            urlcolor=orange}

\newcommand{\I}{{\rm i}}
\renewcommand{\k}{{\mathbf{k}}}
\newcommand{\q}{{\mathbf{q}}}
\newcommand{\p}{{\mathbf{p}}}

\begin{document}

\title{Ferromagnetic instability in itinerant fcc lattice electron systems \\
with higher order 
van Hove singularities: \\ Functional renormalization group study}
\author{P. A. Igoshev}
\affiliation{Institute of Metal Physics, \\620108, Kovalevskaya str. 18, Ekaterinburg, Russia}
\author{A. A. Katanin}
\affiliation{Center for Photonics and 2D Materials, Moscow Institute of Physics and Technology, Institutsky lane 9, Dolgoprudny, 141700 Moscow region, Russia
}
\affiliation{Institute of Metal Physics, \\620108, Kovalevskaya str. 18, Ekaterinburg, Russia}

\begin{abstract}
We investigate the possibility of ferromagnetic ordering in the non-degenerate Hubbard model on the face-centered cubic lattice within the functional renormalization group technique using temperature as a scale parameter. 
We assume the relations between nearest, next-nearest, and next-next-nearest hopping parameters providing higher order (giant) van Hove singularity of the density of states. 
The ferromagnetic instability formation with lowering temperature is described consistently in the one-loop approximation for a one-particle irreducible vertex of two-particle electron interaction. 
The chemical potential versus temperature phase diagrams are calculated. We find ferromagnetic order only for sufficiently strong divergence of the density of states and fillings in the vicinity of van Hove singularity. The obtained Curie temperature is more than an order of magnitude smaller than the results of the random-phase approximation. The main origin of the suppression of ferromagnetism is the screening of interaction in the particle-particle channel. We also do not find the pronounced tendency towards incommensurate order when the Fermi level is moved away from a van Hove singularity, such that the first order quantum phase transitions from the ferro- to paramagnetic phase are obtained. 
\end{abstract}

\maketitle

\section{Introduction}

Weak itinerant magnets, such as ZrZn$_2$~\cite{1958:ZrZn2_discovery,2020:Skornyakov}, Ni$_3$Al~\cite{1984:Ni3Al:Sasakura,2005:Niklowitz}, and some other compounds, represent substances having low magnetic transition temperatures and small saturation magnetic moment. The problem of description of itinerant ferromagnetism is a long-standing problem, which started from the Stoner theory~\cite{Stoner,1966:Penn}. Although this theory describes main features of magnetic transitions, it strongly overestimates magnetic transition temperatures, does not explain the Curie-Weiss behavior of magnetic susceptibilities, and yields an incorrect critical behavior. These effects occur due to electronic and magnetic correlations, which are not accounted by the Stoner theory.

Some drawbacks of the Stoner theory were improved by spin-fluctuation approaches, such as Murata-Doniach~\cite{1972:Murata}, Moriya~\cite{Moriya_book} theory, etc. These approaches account for the effect of correlations by considering the contribution of the paramagnon interaction to the magnetic susceptibility. While Murata-Doniach theory reproduces the Curie-Wess law, it accounts only for the static spin correlations. The dynamic spin correlations were accounted by Moriya theory, which yields the temperature $T$ dependence of magnetic susceptibility $\propto 1/(T^{4/3}-T_{\rm C}^{4/3})$ ($T_{\rm C}$ being the Curie temperature), which is close to the Curie-Weiss law. The Moriya theory, based on the bosonic mean-field approximation, was justified within the Hertz-Millis renormalization group approach \cite{1976:Hertz,1993:Millis} and it was shown that Moriya theory yields qualitatively correct results since the effective dimension $d+z=6$ for the cubic itinerant ferromagnets ($z=3$ is the dynamic critical exponent) is larger than the upper critical dimension $d_c=4$ of the $\phi^4$ theory.

Yet, the spin-fluctuation approaches consider an effect of only a single (particle-hole) channel of electron interaction, which represents the paramagnon contribution. 
At the same time, it was argued that for two-dimensional systems, the singular behavior of the density of states near the Fermi level yields an interplay of different channels of electron interactions~\cite{2001:Katanin,2001:Salmhofer,Kampf}.
In three dimensions, peaks of the density
of states that originate from the so called ``giant'' or ``higher-order'' van Hove singularities (vHs), play the important role of the ferromagnetism formation~\cite{1993:Katsnelson,2019:Igoshev_PMM,2019:Igoshev_JETP,2020:Classen,2022:Igoshev,2022:Betouras}. 
These singularities originate from vanishing of some inverse masses of electronic dispersion at van Hove points, which makes the dispersion effectively two- or one-dimensional. 
A number of band structure investigations point out appearing van Hove singularities of electron density of states, which play the important role for the physics of itinerant magnets. For ZrZn$_2$, the peak of density of states is formed by the vicinity of L point of the Brillouin zone ~\cite{2007:Yamaji,2004:Major,2003:Yates,2020:Skornyakov}, while for Ni$_3$Al, the van Hove point R produces the peak in the density of states~\cite{2011:Hamid}. 
In nickel, which is a stronger metallic magnet, there is a flat band in the vicinity of L point of the Brillouin zone~\cite{2017:Hausoel}. 
For bcc (body centered cubic) iron $e_{\rm g}$ derived band appears to be flat in several direction which, together with Hund coupling $J_{\rm H}$, results in the local moment formation~\cite{OurFe,2015:Efremov}.  

A large enough value of the density of states (DOS) at the Fermi level, occurring due to vHs, leads to the ferromagnetic ground state according to the Stoner theory~\cite{Stoner,1966:Penn}. Therefore, on one hand, the fundamental origin of the itinerant ferromagnetism is the presence of van Hove singularities of itinerant electron electronic dispersion. On the other, the van Hove singularities induce the interplay of different channels of electron scattering and, therefore, require going beyond the spin fluctuation approaches. In particular, the Kanamori screening \cite{Kanamori} is expected to be particularly important near van Hove singularities, see also Refs.~\onlinecite{Tmatrix1,3body1,3body2}. 

In this respect, dynamical mean-field theory provides a possibility for studying the effect of local electronic correlations on ferromagnetism in the single-band \cite{Ulmke,Pruschke} and multiband \cite{Arita,Licht,OurFe,2020:Skornyakov,2017:Hausoel} models. 
In particular, the problem of ferromagnetism formation in the single-band Hubbard model on fcc (face centered cubic) lattices was considered within DMFT [31] in cases (i) $d = \infty$ at the ratio of next-nearest and nearest neighbor hopping parameters $t'/t = 1/2$ (which provides a giant one-dimensional-like van Hove singularity at the bottom of
the band) and (ii) $d = 3$ at $t'=t/4$ which yields a wide and moderate-valued DOS plateau in the vicinity of band bottom (see also Ref.~\onlinecite{2022:Igoshev}).  While in case (i) wide ferromagnetic region was found for Coulomb interaction parameter $U\sim 7t$, in case (ii), ferromagnetism was not found at $U \sim 14t$, but found at $U\sim 21t$. These DMFT results imply that ferromagnetic ordering for the non-degenerate model in the absence of strong peak of density of states can be stable only at very large interaction parameter $U$ values. 
However, the treatment of the non-local effects is beyond the scope of the dynamical mean-field theory. Moreover, a generalization of the Stoner theory to treat the spin spiral instability and N\'{e}el antiferromagnetism reveals also that in a major part of the phase diagram,  the spiral phase is even more preferable than the ferromagnetic one \cite{Schulz,2010:Igoshev,2015:Igoshev}. Therefore the theory of itinerant ferromagnetism should also treat its competition with the other type of magnetic orderings, including incommensurate ones.

A tool allowing to treat both, the interplay of different channels of electron interaction and the competition of ferromagnetic and incommensurate correlations 
is the functional renormalization group (fRG) technique, which is a powerful method for treatment of correlation effects~\cite{
fRG,Zanchi,Halboth,2001:Salmhofer,Kampf,Honerkamp,Rohe,Lauscher,2009:Husemann,Katanin,2013:Eberlein,2017:TUfRG,2022:BFfRG,2011:Igoshev,Yuki,Honerkamp01}.
The fRG directly takes into account non-local electron correlations which are of a great importance in the case of weak itinerant magnetism. 
This approach manifested its success for consideration of the problem of ferromagnetism of two-dimensional Hubbard model due to the closeness of the Fermi level to logarithmic van Hove singularity~\cite{2001:Katanin,Honerkamp,Yuki,Kampf,2011:Igoshev}. 
It was found that necessary condition for ferromagnetism formation is its ``parameteric separation'' from other types of instabilities like superconducting or antiferromagnetic one, which can be achieved by a strong curvature of the Fermi surface caused by~large~$t'/t$, where $t(t')$  is the nearest- (next-nearest-) neighbor hopping parameter (see also Refs. \onlinecite{2007:Igoshev,2009:Igoshev,2010:Igoshev}). The ferromagnetic instability is enhanced in the flat band case, when one of the inverse masses of the electronic dispersion vanishes. This case in three dimensions, corresponding to higher-order van Hove singularity, represents therefore a certain interest for the investigation. 

While the early fRG approaches for fermionic systems were based on the patch parametrization of electron interaction vertices, it was suggested by Husemann and Salmhofer \cite{2009:Husemann} to parametrize the flow of each interaction channel by the bosonic momentum (and possibly frequency) transfer, while the remaining dependence on two fermionic momenta and frequencies can be accounted less accurately via their projections onto some restricted set of basis functions. This idea was further developed within the truncated unity functional renormalization group approach \cite{2017:TUfRG}, which suggests projection of the ``transverse" channels onto the same set of basis functions, which greatly simplifies calculations. Finally, recently the formulation of the approach of Husemann and Salmhofer in terms of the triangular boson-fermion vertices and the reminder was proposed \cite{2022:BFfRG}.  

These recent developments give a possibility to study itinerant ferromagnetism of the three-dimensional fermionic systems with sufficiently small number of flowing vertices.
In the present paper, we investigate the formation of ferromagnetism on fcc lattice in the presence of extended van Hove singularities. In particular, we study (i)~how the strength of the van Hove singularity affects stability of the ferromagnetism;
(ii)~what is the main destructive factor of ferromagnetism in three dimensions: a competition with incommensurate magnetic order or the competition with the paramagnetic phase, and~(iii)~how important is the effect of screening of Coulomb interaction.

To study the above mentioned problems, we present a study of the evolution of magnetic and electronic properties with decreasing temperature within 1PI fRG for the non-degenerate Hubbard model on the fcc lattice. The plan of the paper is the following.
In Sec.~\ref{sec:model_diagram}, the model and fRG equations are presented. 
In Sec.~\ref{sec:results}, we present and discuss the numerical results. In Sec.~\ref{sec:conclusions}, we conclude the paper. 
\vspace{-0.1cm}
\section{The model and RG equations}\label{sec:model_diagram}
\subsection{The model}
We consider the Hubbard model with the Hamiltonian 
\begin{equation}
\mathcal{H}=\sum_{{\bf k}\sigma }\epsilon_{\mathbf{k}%
}c_{{\bf k}\sigma }^{+}c_{{\bf k}\sigma }+\frac{U}{2N}\sum_{{\bf k}_{1}{\bf k}_{2}{\bf k}_{3}{\bf k}_{4},\sigma \sigma ^{\prime }}c_{\mathbf{k}_{1}\sigma
}^{+}c_{\mathbf{k}_{2}\sigma ^{\prime }}^{+}c_{\mathbf{k}_{3}\sigma ^{\prime }}c_{\mathbf{k}_{4}\sigma
}\delta_{{\bf k}_{1}+\mathbf{k}_{2},\mathbf{k}_{3}+\mathbf{k}_{4}},  
\label{Action}
\end{equation}%
where $\epsilon_{\mathbf{k}}$ is an electronic dispersion, $U$ is the Hubbard on--site interaction, $N$ is lattice site number, $\delta$ is the Kronecker $\delta$-symbol, $c_{{\bf k}\sigma}$ and $c^+_{{\bf k}\sigma}$ are the electron destruction and creation operators, respectively. We consider electronic dispersion on the face-centered cubic lattice  with nearest- (unity), next-nearest $\tau$, and next-next-nearest $\tau'$ hoppings:
\begin{multline}\label{eq:bare_spectrum}
\epsilon_{\k}(\tau,\tau') = 
-4\left(\cos\frac{k_x}2\cos\frac{k_y}2 +\cos\frac{k_x}2\cos\frac{k_z}2 + \cos\frac{k_y}2\cos\frac{k_z}2\right)
\\
+ 
\vphantom{\frac{k_x}2}
2\tau\left(\cos k_x + \cos k_y + \cos k_z\right)
+
4\tau'\left(\cos k_x\cos k_y +\cos k_x\cos k_z + \cos k_y\cos k_z\right),
\end{multline}
{where the lattice constant is put to unity}.

\subsection{van Hove singularities\label{Sect:vH}} 
We are interested in studying itinerant magnetism induced by higher order van Hove singularities. 
\begin{figure}[b]
\includegraphics[angle=-90,width=.5\textwidth]{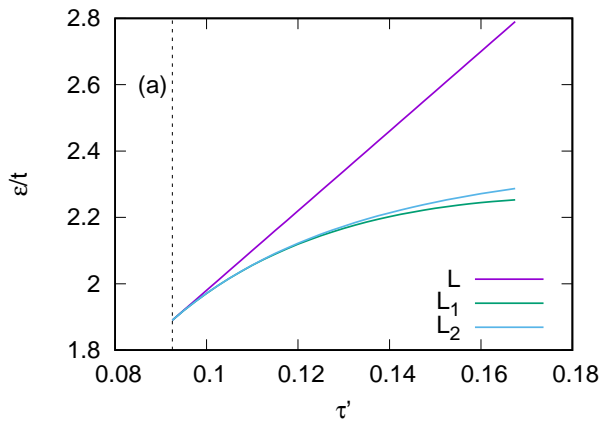}
\caption{$\tau'$ dependence of van Hove energies at $\tau = -0.13$.}
\label{fig:vHS}
\end{figure}
For itinerant ferromagnets on fcc lattice peculiarities of the dispersion in the vicinity of L$(\pi,\pi,\pi)$ point represent certain interest. 
To investigate the contribution of the vicinity of L$(\pi,\pi,\pi)$ point to DOS, we expand the~Eq.~(\ref{eq:bare_spectrum}) as $\epsilon_{\k}(\tau,\tau') = \epsilon_{\k_{\rm L}}(\tau,\tau') +(\tau - 4\tau')q^2 - q_yq_z - q_zq_x - q_xq_y$, where $\k = \k_{\rm L} + \q$ and $\epsilon_{\k_{\rm L}}(\tau,\tau') = 6(-\tau + 2\tau')$. Since the eigenvalues of this $\q$-quadratic form (inverse masses) are $m^{-1}_{\rm L,1} = 1 + 2\tau - 8\tau'$ (doubly degenerate) and $m^{-1}_{\rm L,2} = 2(-1 + \tau - 4\tau')$, we obtain the condition on giant (quasi-one-dimensional-like) van~Hove singularity at
\begin{equation}\label{eq:giant_vHS_condition}
\tau'_{\rm c}(\tau) = \frac18 + \frac{\tau}4. 
\end{equation}

\begin{figure}[b]
\includegraphics[angle=-90,width=.49\textwidth]{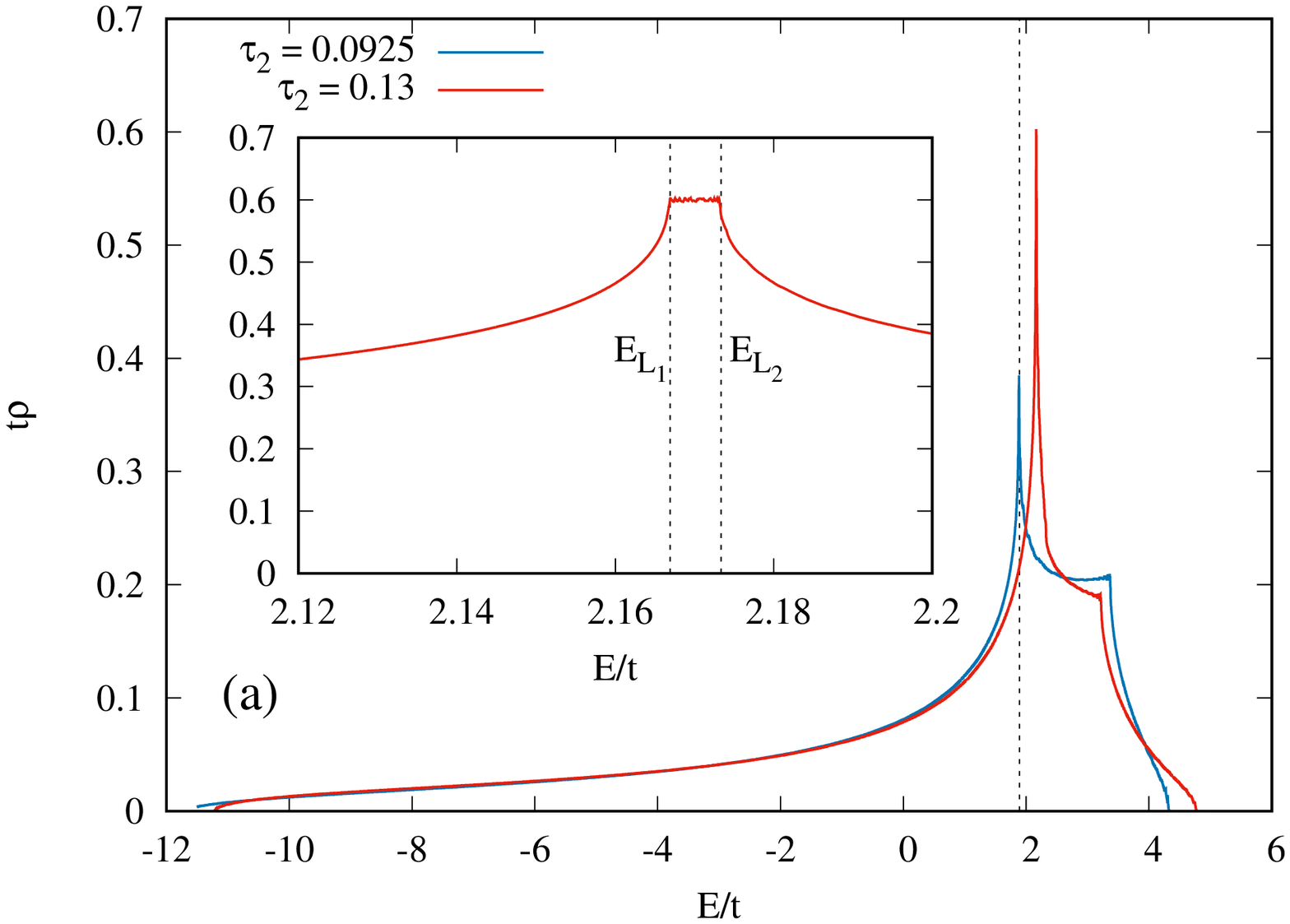}
\includegraphics[angle=-90,width=.49\textwidth]{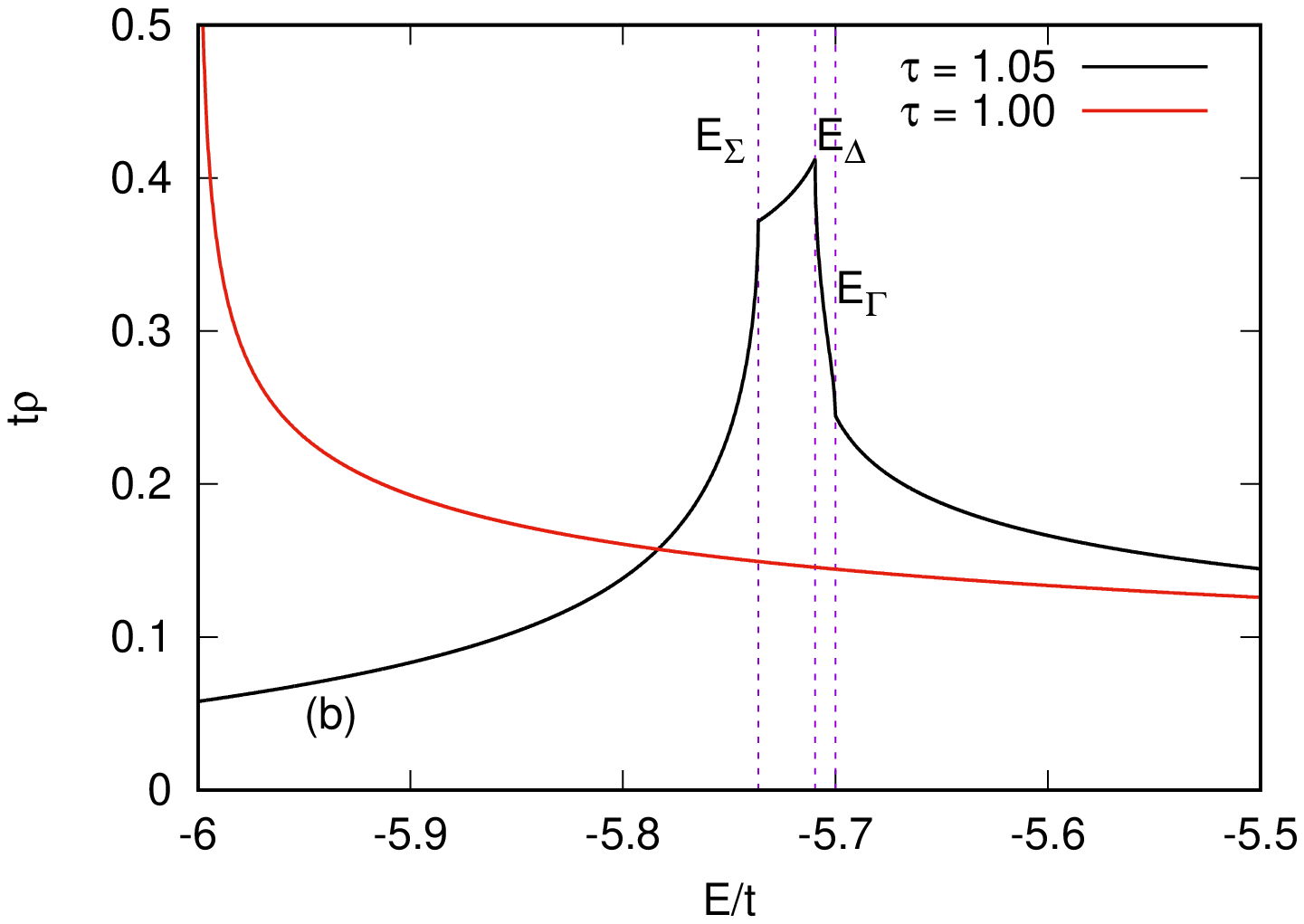}
\caption{(a) Density of states $\rho(\epsilon,\tau,\tau')$ plot as a function of $\epsilon$ for $\tau = -0.13$, $\tau' = 0.0925$ and $\tau'=0.13$. In the inset, $\rho(\epsilon,\tau,\tau')$ at $\tau' = 0.13$ in the energy in the window near the van Hove structure, corresponding to the energies $E_{\rm L_1}$ and  $E_{\rm L_2}$ (marked by dashed vertical lines) is shown. (b) The density of states in the vicinity of the bottom of the band for $\tau' = 0$, $\tau = 1$~(red solid line) and $\tau = 1.05$~(black solid line). For the case $\tau = 1.05$, the positions of van Hove structure energies~\cite{2022:Igoshev} $E_{\varSigma^\ast} = -5.736$, $E_{\varDelta^\ast} = -5.71$ and $E_{\Gamma} = -5.7$ are shown by vertical dashed lines. 
}
\label{fig:dos}
\end{figure}

The investigation of the electronic dispersion in the vicinity of L~point at $\tau'\ne\tau'_{\rm c}(\tau)$ (see Appendix A) yields the following topological transition in the profile of isoenergy surfaces: While at $\tau' < \tau'_{\rm c}(\tau)$ we have single van Hove point L with finite masses, at $\tau'>\tau'_{\rm c}(\tau)$ the van Hove point L splits into two van Hove points L$_1=(\pi - \delta k_1, \pi + \delta k_1, \pi)$ (on the face of the Brillouin zone) and L$_2=(\pi - \delta k_2, \pi - \delta k_2', \pi - \delta k_2')$, where
$\delta k_1 \simeq 2\sqrt{\delta \tau'/\tau'}$, 
$\delta k_2(\tau, \tau')  \approx -4\sqrt{{\delta\tau'}/({3\tau'})},$ 
$\delta k_2'(\tau, \tau') \approx 2\sqrt{{\delta\tau'}/({3\tau'})}$ at small $\delta\tau'= \tau' - \tau'_{\rm c}$. The energies of these van Hove points as functions of $\tau'$ at $\tau = -0.13$ are shown in~Fig.~\ref{fig:vHS}. 
One can see that despite the fact that $L_{1,2}$ points are well separated in reciprocal space, their energies $E_{L_{1,2}}$  are very close and their difference increases only slowly with increasing of $\tau'$.  This scenario of high-order van Hove point splitting was earlier found for cubic lattices with the dispersion, containing only nearest- and next-nearest neighbor  hoppings~\cite{2019:Igoshev_PMM,2019:Igoshev_JETP,2022:Igoshev}.
Using the tetrahedron method~\cite{1994:Andersen}, we calculate numerically the density of states 
$\rho(\epsilon, \tau, \tau') = \sum_{\k}\delta(\epsilon - \epsilon_{\k}(\tau,\tau'))$,
see Fig.~\ref{fig:dos}(a). {Here and below we include the factor $1/N$ into the definition of the sum over Brillouin zone.} One can see that the  plateau with finite but rather high value of density of states occurs in the energy interval $\left[E_{\rm L_1}, E_{\rm L_2}\right]$. 
At~$\tau' = \tau'_{\rm c}(\tau)$, where $\tau'_{\rm c}(\tau)$ is determined by the Eq.~(\ref{eq:giant_vHS_condition}), the~DOS has a~logarithmic van Hove singularity:  $\rho(\epsilon,\tau,\tau'_{\rm c}(\tau))\sim {8}/({\pi^2{{\sqrt{3(1+2\tau)}}}})\ln\left({4}/{|\epsilon-\epsilon_{\mathbf{k}_{\rm L}}(\tau,\tau'_{\rm c})|}\right)$ (see Appendix~\ref{sec:spectrum_and_DOS}).

We also compare the above discussed van Hove singularities to the case of high-order van Hove singularity for the dispersion on the fcc lattice with only nearest and next-nearest hopping ($\tau ' = 0$)~\cite{2022:Igoshev}. In this case, the~high-order van Hove point is the $\Gamma$ point of the Brillouin zone;  the van Hove singularity occurs when $\tau = 1$ and it is located at the bottom of the band, $\rho(\epsilon, 1, 0)\propto (\epsilon + 6)^{-1/4}$, see Fig.~\ref{fig:dos}(b). 
While $\tau$ deviates from 1, the van Hove point splits onto the van Hove structure corresponding to the points ${\varSigma^\ast}$ and ${\varDelta^\ast}$, which lie on high-symmetry directions $\varSigma(p,p,0)$ and $\varDelta(p,0,0)$ in the vicinity of the $\Gamma$ point and have close energies provided that $\tau - 1\ll 1$. While the energy interval $[E_{\varSigma^*},E_{\varDelta^*}]$ does not formally correspond to divergence of DOS, it can provide rather high value of DOS, 
which is of a great importance for itinerant ferromagnetism stability. See an example of DOS in Fig.~\ref{fig:dos}(b) at $\tau = 1.05$ and more details in Ref.~\onlinecite{2022:Igoshev}.

\subsection{fRG equations}\label{sec:fRG_equations}
We use the 1PI fRG equations~\cite{fRG} for 1PI 2-particle vertex, which in the absence of magnetic field and magnetic order (i.e., for the SU(2) symmetry) has the~form
\begin{equation}
\Gamma_{\sigma_1\sigma_2;\sigma_3\sigma_4}(p_1,p_2,p_3) = V(p_1,p_2,p_3)\delta_{\sigma_1\sigma_3}\delta_{\sigma_2\sigma_4} - V(p_1,p_2,p_4)\delta_{\sigma_1\sigma_4}\delta_{\sigma_2\sigma_3}.
\end{equation}
The momentum dependence of the vertex $V$ is of particular interest. To parametrize this, we introduce following C.~Husemann and M.~Salmhofer~\cite{2009:Husemann} bosonic ($l$) and fermionic ($k_{1,2}$) momenta in various channels:
\begin{eqnarray}
\label{eq:SC_momentum_def}
    l^{\rm P} &=& p_1 + p_2, k^{\rm P}_1 = (p_1 - p_2)/2, k^{\rm P}_2 = (p_4 - p_3)/2, \\
\label{eq:M_momentum_def}
    l^{\rm M} &=& p_1 - p_3, k^{\rm M}_1 = (p_1 + p_3)/2, k^{\rm M}_2 = (p_2 + p_4)/2, \\
\label{eq:K_momentum_def}
    l^{\rm K} &=& p_3 - p_2, k^{\rm K}_1 = (p_1 + p_4)/2, k^{\rm K}_2 = (p_2 + p_3)/2.
\end{eqnarray}
Within the channel decoupling representation
\begin{align}\label{eq:V}
        V(p_1,p_2,p_3) &= U -\Phi_{\rm SC}(l^{\rm P}, k_1^{\rm P},k_2^{\rm P}) 
        + \Phi_{\rm M}(l^{\rm M}, k_1^{\rm M}, k_2^{\rm M}) \notag\\
        &+ \dfrac{1}{2}\Phi_{\rm M}(l^{\rm K}, k_1^{\rm K}, k_2^{\rm K}) - \dfrac{1}{2}\Phi_{\rm K}(l^{\rm K}, k_1^{\rm K}, k_2^{\rm K}),
\end{align}
we have the following equations for constituents of the vertex~$V$
\begin{eqnarray}
    \label{eq:main_SC}
    \dot{\Phi}_{\rm SC}(l^{\rm P}, k_1^{\rm P}, k_2^{\rm P}) &=& -\tau_{\rm pp}(p_1,p_2,p_3),\\
    \label{eq:main_M}
    \dot{\Phi}_{\rm M}(l^{\rm M}, k_1^{\rm M}, k_2^{\rm M}) &=& \tau_{\rm ph}^{\rm cr}(p_1,p_2,p_3),\\
    \label{eq:main_K}
    \dot{\Phi}_{\rm K}(l^{\rm K}, k_1^{\rm K}, k_2^{\rm K}) &=& -2\tau_{\rm ph}^{\rm d}(p_1,p_2,p_3) + \tau_{\rm ph}^{\rm cr}(p_1,p_2,p_4),
\end{eqnarray}
where the derivative is taken with respect to the flow parameter $s = \ln(t/T)$ and explicit expressions for $\tau_{\rm pp}(p_1,p_2,p_3)$, $\tau_{\rm ph}^{\rm d}(p_1,p_2,p_3)$, and $\tau_{\rm ph}^{\rm cr}(p_1,p_2,p_3)$ are given in Appendix~\ref{AppB}. 

In the following, we neglect the frequency dependencies of the vertices, which is justified in the weak-coupling regime. To treat the momentum dependencies, we apply truncated unity fRG approach \cite{2017:TUfRG}. To this end, we define the projection operators acting on functions of the momenta $F(p_1,p_2,p_3)$ as
\begin{eqnarray}
    \label{eq:SC_projection}
    F^{\rm P}_{n,m}(\mathbf{l}) &=& \sum_{\k\k'} f_n(\textbf{k}) f_m(\textbf{k}') F\left(\dfrac{\mathbf{l}}{2} + \textbf{k},\dfrac{\mathbf{l}}{2} - \textbf{k},\dfrac{\mathbf{l}}{2} - \textbf{k}'\right),\\
    \label{eq:M_projection}
    F^{\rm M}_{n,m}(\mathbf{l}) &=& \sum_{\k\k'} f_n(\textbf{k}) f_m(\textbf{k}') F\left(\dfrac{\mathbf{l}}{2} + \textbf{k},{\mathbf k}' - \dfrac{\mathbf{l}}{2},{\mathbf k} - \dfrac{\mathbf{l}}{2}\right),\\
    \label{eq:K_projection}
    F^{\rm K}_{n,m}(\mathbf{l}) &=& \sum_{\k\k'} f_n(\textbf{k}) f_m(\textbf{k}') F\left(\dfrac{\mathbf{l}}{2} + \textbf{k},\textbf{k}' - \dfrac{\mathbf{l}}{2},\dfrac{\mathbf{l}}{2} + \textbf{k}'\right),
\end{eqnarray}
where a set of functions $f_n(\k)$ constitute a basis in the corresponding Hilbert space.  
We apply these operators to Eqs.~(\ref{eq:main_SC})--(\ref{eq:main_K}). Using the projection rules 
(\ref{eq:SC_projection}) and (\ref{eq:M_projection}), and inserting representation of unity in the right-hand sides (r.h.s.), we obtain the projected components of the r.h.s. (see Appendix \ref{AppB})
\begin{eqnarray}
    (\tau_{\rm pp})^{\rm P}_{n_1,n_2}(\mathbf{l}) &= \sum_{m_1,m_2}V^{\rm P}_{n_1,m_1}(\mathbf{l})\chi^{\rm pp}_{m_1,m_2}(\mathbf{l})V^{\rm P}_{m_2,n_2}(\mathbf{l}),\\
    (\tau^{\rm cr}_{\rm ph})^{\rm M}_{n_1,n_2}(\mathbf{l}) &= \sum_{m_1,m_2}V^{\rm M}_{n_1,m_1}(\mathbf{l})\chi^{\rm ph}_{m_1,m_2}(\mathbf{l})V^{\rm M}_{m_2,n_2}(\mathbf{l}),
\end{eqnarray}
where the explicit form of $V^{\rm P,M}(\mathbf{l})$ is given by the Eqs.~(\ref{eq:SC_projection}) and (\ref{eq:M_projection}) with $F = V$,
\begin{align}\label{eq:chi_pp_def}
   \chi^{\rm pp,ph}_{n,m}(\mathbf{l}) &= \sum_\mathbf{p} f_n(\mathbf{p}) f_m(\mathbf{p})\sum_{\nu}L_{\rm pp,ph}(\mathbf{l},p),\\
    L_{\rm pp,ph}(\mathbf{l}, p) &= -\partial_{s}\left[G(\mathbf{l}/2 + {\bf p},\nu)G(\pm \mathbf{l}/2 \mp {\bf p},\mp \nu)\right],
\end{align}
$G({\bf p},\nu)$ is the electron Green's function. 
Finally,
\begin{multline}
 (\tau^{\rm d}_{\rm ph})^{\rm K}_{n_1,n_2}(\mathbf{l}) = -2\sum_{m_1,m_2}V^{\rm K}_{n_1,m_1}(\mathbf{l})\chi^{\rm ph}_{m_1,m_2}(\mathbf{l})V^{\rm K}_{m_2,n_2}(\mathbf{l}) 
 \\
 + \sum_{m_1,m_2}[V_{\rm ex}]^{\rm K}_{n_1,m_1}(\mathbf{l})\chi^{\rm ph}_{m_1,m_2}(\mathbf{l})V^{\rm K}_{m_2,n_2}(\mathbf{l}) 
 + \sum_{m_1,m_2}V^{\rm K}_{n_1,m_1}(\mathbf{l})\chi^{\rm ph}_{m_1,m_2}(\mathbf{l})[V_{\rm ex}]^{\rm K}_{m_2,n_2}(\mathbf{l}),
\end{multline}
where the index ``ex'' corresponds to an interchange of third and forth 4-vertex argument:
$V_{\rm ex}(\mathbf{l}/2 + \mathbf{k}_1, \mathbf{k}_2 - \mathbf{l}/2, \mathbf{l}/2 + \mathbf{k}_2) = V(\mathbf{l}/2 + \mathbf{k}_1, \mathbf{k}_2 - \mathbf{l}/2, \mathbf{k}_1 - \mathbf{l}/2)$,
and
\begin{equation}
    (\tau^{\rm cr}_{\rm ph, ex})^{\rm K}_{n_1,n_2}(\mathbf{l}) = \sum_{m_1,m_2}V^{\rm M}_{n_1,m_1}(\mathbf{l})\chi^{\rm ph}_{m_1,m_2}(\mathbf{l})V^{\rm M}_{m_2,n_2}(\mathbf{l}),
\end{equation}
the explicit form of $V^{\rm K}(\mathbf{l})$ is given by the Eq.~(\ref{eq:K_projection}) with $F = V$ and we have used 
    ${F}^{\rm K}_{\rm ex} = {F}^{\rm M}$.

Denoting matrices with respect to the projection indexes by bold capital letters, we rewrite Eqs.~(\ref{eq:main_SC})--(\ref{eq:main_K}) in the matrix form in terms of projections:
\begin{eqnarray}
    \label{eq:Phi_SC_final}
    \dot{\bm{\Phi}}^{\rm SC}(\mathbf{l}) &=& -\mathbf{V}^{\rm P}(\mathbf{l})\cdot{\bm{\chi}}^{\rm pp}(\mathbf{l})\cdot\mathbf{V}^{\rm P}(\mathbf{l}), \\
    \label{eq:Phi_M_final}
    \dot{\bm{\Phi}}^{\rm M}(\mathbf{l}) &=& \mathbf{V}^{\rm M}(\mathbf{l})\cdot{\bm{\chi}}^{\rm ph}(\mathbf{l})\cdot\mathbf{V}^{\rm M}(\mathbf{l}), \\ 
    \label{eq:Phi_K_final}
    \dot{\bm{\Phi}}^{\rm K}(\mathbf{l}) &=& (2\mathbf{V}^{\rm K}(\mathbf{l}) - \mathbf{V}^{\rm M}(\mathbf{l}))\cdot{\bm{\chi}}^{\rm ph}(\mathbf{l})\cdot(2\mathbf{V}^{\rm K}(\mathbf{l}) - \mathbf{V}^{\rm M}(\mathbf{l})),
\end{eqnarray}
where $\cdot$ means matrix multiplication. 
We consider the temperature flow\cite{2001:Salmhofer} with $s = -\ln T$, $G({\bf p},\nu) = {T^{1/2}}/({\I\nu - \xi_{\mathbf{p}}})$, and 
\begin{equation}
    \sum_\nu L_{\rm pp,ph}({\bf l},p) =
\pm\partial_s\frac{f[\xi_{\mathbf{p} + \mathbf{l}/2}] - f[\mp\xi_{\mathbf{p} - \mathbf{l}/2}]}{\xi_{\mathbf{p} + \mathbf{l}/2} \pm \xi_{\mathbf{p} - \mathbf{l}/2}}.
\end{equation}
Here $\xi_{\mathbf{p}}= \varepsilon_{{\mathbf p}} - \mu$ ($\mu$ being the chemical potential), $f[\xi] = 1/(\exp(\xi/T)+1)$ is the Fermi function. 
Since $\partial_s{f}[\xi] = \xi f'[\xi]$, we get
\begin{equation}
   \chi^{\rm pp,ph}_{n,m}(\mathbf{l}) = \pm\sum_{\mathbf{p}} f_n(\mathbf{p}) f_m(\mathbf{p})\frac{\xi_{\mathbf{p} + \mathbf{l}/2} f'[\xi_{\mathbf{p} + \mathbf{l}/2}]  \pm\xi_{\mathbf{p} - \mathbf{l}/2}f'[\mp\xi_{\mathbf{p} - \mathbf{l}/2}]}{\xi_{\mathbf{p} + \mathbf{l}/2} \pm \xi_{\mathbf{p} - \mathbf{l}/2}},
\end{equation}


We restrict consideration to the trivial set of the ($s$-wave) basis functions $f_m={\rm const}$, which corresponds to momentum averaging of the vertex taken in the ``transverse" channels. The fRG equations can be written in the following explicit form  within the channel decoupling scheme
\begin{eqnarray}
\label{eq:dot_Phi_SC}
	\dot\Phi_{\rm SC}(\q) &=& -V_{\rm SC}^2(\q) L_{+}(\q),\\
 \label{eq:dot_Phi_M}
	\dot\Phi_{\rm M}(\q) &=& +V_{\rm M}^2(\q) L_{-}(\q),\\
 \label{eq:dot_Phi_K}
	\dot\Phi_{\rm K}(\q) &=& +V_{\rm K}^2(\q)L_{-}(\q),
\end{eqnarray}
where 
\begin{eqnarray}
	V_{\rm SC}(\q) &=& U - \Phi_{\rm SC}(\q) + \frac32\bar\Phi_{\rm M} - \frac12\bar\Phi_{\rm K},\\
	V_{\rm M}(\q) &=& U + \Phi_{\rm M}(\q) - \bar{\Phi}_{\rm SC} + \frac12\bar{\Phi}_{\rm M} - \frac12\bar{\Phi}_{\rm K},\\
	V_{\rm K}(\q) &=& U - \Phi_{\rm K}(\q) - \bar{\Phi}_{\rm SC} + \frac32\bar{\Phi}_{\rm M} + \frac12\bar{\Phi}_{\rm K},
\end{eqnarray}
\begin{equation}\label{eq:L_def}
L_{\pm}(\q) = \pm\sum_{\p}\frac{\xi_{\p + \q/2}f'_{\p + \q/2} \pm
	\xi_{\p - \q/2}f'_{\p - \q/2}}{\xi_{\p + \q/2} \pm \xi_{\p - \q/2}},
\end{equation}
and 
\begin{equation}
	\bar{\Phi}_c \equiv \sum_{\q} \Phi_c(\q),\; c = \text{SC, M, K}.
\end{equation}

The fRG equations (\ref{eq:dot_Phi_SC})--(\ref{eq:dot_Phi_K}) form the system of ordinary \textit{functional} differential equations. These equations also allow single-boson exchange decomposition \cite{2022:BFfRG}, see Appendix \ref{AppC}. Below we present the procedure of numerical solution of the equations (\ref{eq:dot_Phi_SC})--(\ref{eq:dot_Phi_K}) in details. 

\subsection{Numerical implementation}

Since the polarization functions derivatives $L_{\pm}({\bf q})$ are independent of $\Phi_{\rm SC,M,K}(\q)$, we calculate the scale $s$ dependence of $L_{\pm}({\bf q})$ on a grid of $s$ and $\q$. The $s$ dependence of $L_{\pm}({\bf q})$ is further interpolated by splines. We use the $\q$ grid in the Brillouin zone with the number of reciprocal vector division $n_{\q} = 8$, which corresponds to 29 different points in the part of the Brillouin zone, irreducible by the point group symmetry. 
We have verified that the results do not change qualitatively for $n_{\q}=16$. 
To perform integration over $\k$ in Eq. (\ref{eq:L_def}) we consider adaptive division of each of the $6n^3_{\q}$ tetrahedrons in the standard uniform grid of the Brillouin zone using TOMS algorithm~\cite{1993:Berntsen}.
Here we take into in account the fact that $\Phi_{\rm SC,M,K}(\mathbf{q})$ is a more smooth function of $\mathbf{q}$, than the electronic dispersion $\xi_{\mathbf{k}}$, which details are of a great importance and require using very dense (adaptive) $\mathbf{k}$ grid for integration~(see~Eq.~(\ref{eq:L_def})). 
The obtained functions $L_{\pm}({\bf q})$ 
are used for the solution of ordinary differential equations by Runge-Kutta procedure for different values of $U$ without recalculation~of~$L_{\pm}$. 
The divergence of $\Phi_{\rm M}(\mathbf{Q})$ at some scale $s = s_0$ corresponds to the phase transition into the ordered phase with the wave vector $\mathbf{Q}$ and the corresponding critical temperature (Curie temperature $T_{\rm C}=\exp(-s_0)$ at $\mathbf{Q} = 0$).

\begin{figure}[t]
\includegraphics[angle=-90,width=.65\textwidth]{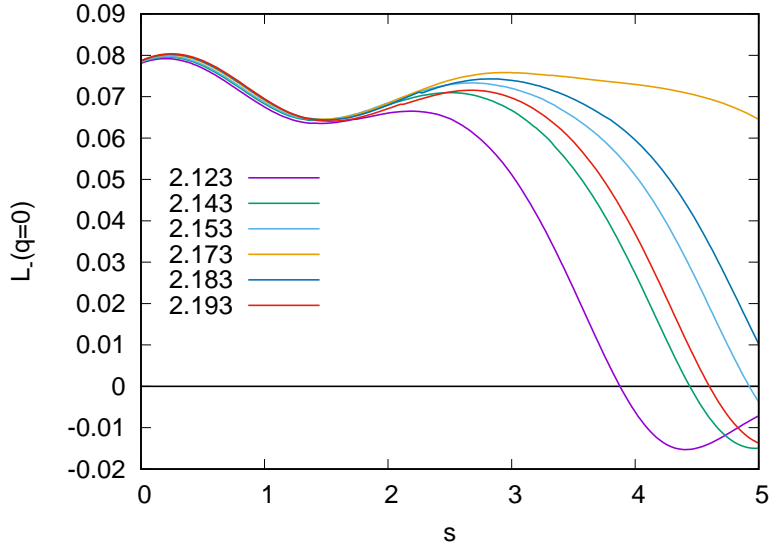}%
\caption{The $s$ dependence of $L_-(\mathbf{q}=0)$, see definition (\ref{eq:L_def}), at $\tau= -\tau'=-0.13$, and different values of $\mu$. Numbers in the legend denote $\mu$.
}
\label{fig:Lph_t1=-0.13_t2=+0.13}
\end{figure}
The $s$ dependence of quantities $L_{\pm}(\mathbf{q})$ determines the fRG flow for specified $U$.  
The most important is $L_{-}(\mathbf{q} = 0)$ dependence, see the examples of its scale dependence in Fig.~\ref{fig:Lph_t1=-0.13_t2=+0.13}. Non-monotonic $s$-dependence of $L_{-}(0)$ originates from the presence of several energy scales of DOS. The first pair of extema  ($0.5 \lesssim s \lesssim 1.5$) corresponds to large scale feature of DOS at $2.5 \lesssim \epsilon \lesssim 3.5$ and its is not sensitive to the variation of the Fermi level position. The second pair of extema (at $3 \lesssim s \lesssim 5$), which is present for $\mu=2.123$, is sensitive to the Fermi level position and originates from the small scale feature (van Hove plateau) of density of states at $2.167 \lesssim \epsilon \lesssim 2.172$.
One can see that for $\mu$ far away from van Hove plateau, $L_{-}(\mathbf{q} = 0)$ becomes negative below some $\mu$-dependent temperature. Therefore below this temperature the formation of ferromagnetic instability becomes impossible, see~Eq.~(\ref{eq:dot_Phi_M}). 
When for fixed $U$ and the temperature $T_{\rm C}(\mu, U)$ the condition $L_{-}(\mathbf{q} = 0) = 0$ is fulfilled, $T_{\rm C}$ jumps to zero which indicates the first order quantum phase transition.

Since the numerical complexity of calculation of $L_{\pm}(\mathbf{Q})$ grows with decreasing temperature due to narrowing of actual integration region domain, only $s \le s_{\rm max}\simeq 6$ region is~accessible.  
For calculation of the phase diagrams we calculate fRG flows at several values of the chemical potential $\mu$ and the dense grid of $U$, determining Curie temperature $T_{\rm C}(\mu, U)$, if it exists. 
Generally, the flow is stopped when the vertices become large, which corresponds to magnetic instability, 
or maximal available scale $s_{\rm max}$ is achieved. 
If the ferromagnetic component of the vertex $\Phi_{\rm M}(\mathbf{q} = 0)$ is strongly enhanced at the end of the flow, the extrapolation procedure is employed for $\Phi_{\rm M}(\mathbf{q} = 0)$ to determine $T_{\rm C}$ beyond the available temperature region.  

\section{Results}\label{sec:results}

In this section we present and discuss numerical results of the present fRG approach for different position of the Fermi level and Coulomb interaction $U$. 

\begin{figure}[h!]
\includegraphics[angle=-90,width=.65\textwidth]{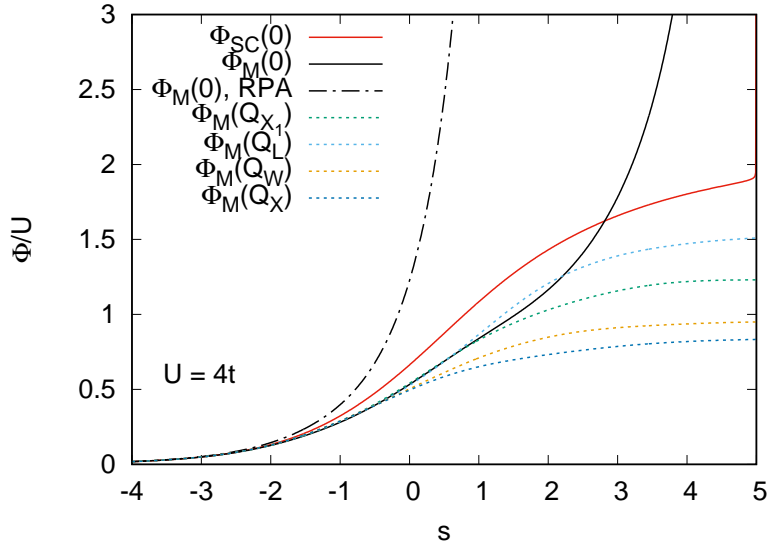}
\caption{Scale (temperature) dependence for components of the vertex $\Phi_{\rm SC}(\mathbf{0})$ and $\Phi_{\rm M}(\mathbf{q})$  in units of $U$ for various symmetric $\q$ vectors for $\tau = -\tau'=-0.13$, $\mu =E_{\rm L_2}=2.17335$, 
$U = 4t$. Double-dot-dashed line shows the RPA result. The Fermi level corresponds to right edge of van Hove plateau, see Fig.~\ref{fig:dos}(b). 
The divergence of $\Phi_M(\mathbf{Q} = 0)$ corresponds to~$s = s_{\rm C} = 5.0$. 
}
\label{fig:fRG_flows_right_edge}
\end{figure}

We consider first the flow for $\tau = -\tau'=-0.13$. For the Fermi level $\mu = E_{L_2}=2.17335t$, corresponding to the right edge of van Hove plateau (see Fig.~\ref{fig:dos}), the fRG flow of the vertices $\Phi_{\rm SC}(0)$ and $\Phi_{\rm M}({\bf Q})$ for $U = 4t$ and symmetric wave vectors $\mathbf{Q} = 0, \mathbf{Q}_{\rm X}=(2\pi,0,0), \mathbf{Q}_{\rm L}=(\pi,\pi,\pi), \mathbf{Q}_{\rm W}=(2\pi,\pi,0), \mathbf{Q}_{{\rm X}_1} = \mathbf{Q}_{\rm X}/2$  is~shown in Fig.~\ref{fig:fRG_flows_right_edge}. The critical temperature of ferromagnetic instability $T_{\rm C}$ manifests itself by the diverging $\Phi_{\rm M}(\mathbf{Q} = 0)$ component of the vertex at $s = s_{\rm C}$. 
The substantial screening particle-particle component of the vertex, $\Phi_{\rm SC}(\mathbf{Q} = 0)\approx (1.5-2)U$ is found in the end of the flow. 
We actually do not find any noticeable incommensurate or antiferromagnetic fluctuation to be enhanced and competing with ferromagnetism formation. In particular, from Fig.~\ref{fig:fRG_flows_right_edge} one can see that  $\Phi_{\rm M}(\mathbf{q})$ is not enhanced remarkable when $\mathbf{q}\ne0$ for the chosen $\mathbf{q}$ points; we have also verified that this holds in the entire Brillouin zone. Therefore the ferromagnetic ordering is suppressed due $\bar\Phi$ contributions into Eq.~(\ref{eq:Phi_M_final}), but not due to competition with incommensurate $\Phi(\mathbf{q}\ne0)$, in contrast to the mean-field studies~\cite{2010:Igoshev,2015:Igoshev}. 
One can also see that the obtained Curie temperature $T_{\rm C}$ is much lower than the corresponding temperature in the random phase approximation (RPA) due to screening of the interaction in the particle-particle channel.

\begin{figure}[t]
\includegraphics[angle=-90,width=.65\textwidth]{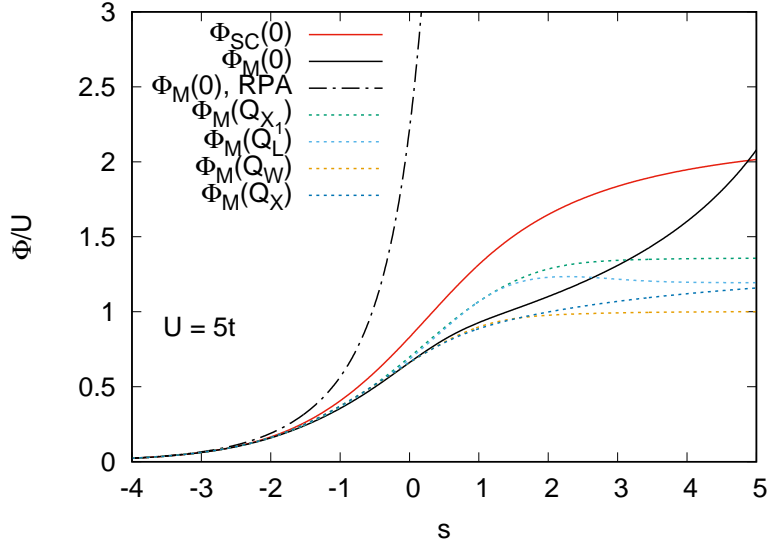}
\caption{$\tau = -0.13$, $\tau'=+0.0925$, $\mu = 1.89$. Scale (temperature) dependence for components of 4-vertex $\Phi_{\rm SC}(\mathbf{Q}=0)$, $\Phi_{\rm M}(\mathbf{q})$ in units of $U$ for symmetric vectors ${\bf q}$, $U = 5t$. Double-dot-dashed line shows the RPA result. The Fermi level corresponds to van Hove singularity at $\mu = 1.89t$, see Fig.~\ref{fig:dos}(a).  
}
\label{fig:fRG_flows_vHS1}
\end{figure}

To show the necessity of having a pair of van Hove points, we contrast the above discussed case $\tau' = 0.13$ (a pair of giant van Hove singularities yielding plateau of of DOS) to the case $\tau' = \tau'_{\rm c}=0.0925$ (weak van Hove singularity $\rho(\epsilon,\tau,\tau'_{\rm c}(\tau))\sim {8}/({\pi^2{{\sqrt{3(1+2\tau)}}}})\ln\left({4}/{|\epsilon-\epsilon_{\mathbf{k}_{\rm L}}(\tau,\tau'_{\rm c})|}\right)$, see Sect. \ref{Sect:vH}). One can see from~Fig.~\ref{fig:fRG_flows_vHS1} that at~$U = 
5t$ no divergence of $\Phi_{\rm M}(\mathbf{Q} = 0)$ is achieved. In this case, the divergence is suppressed by faster growth of the screening component $\Phi_{\rm SC}(\mathbf{Q} = 0)$, which also reaches considerable value 
$\sim (1.5-2)U$. 
The role of incommensurate or antiferromagnetic correlations is the same as in the case $\tau = -\tau'=-0.13$, none of them compete with the ferromagnetic instability. Therefore the plateau of the density of states, or, at least, stronger than logarithmic divergence of the density of states, seems to be the necessary condition for ferromagnetism formation in three dimensions in the single-band Hubbard model.  This conclusion agrees with the results of Ref. \onlinecite{Ulmke}.

\begin{figure}[t]
\includegraphics[angle=-90,width=.65\textwidth]{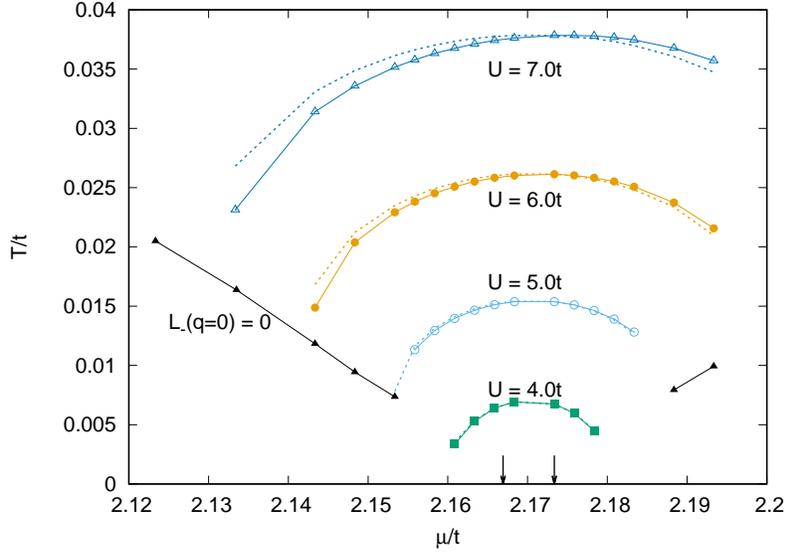}%
\caption{Phase diagram $\mu$-$T_{\rm C}$ for $\tau= -\tau'=-0.13$ at different $U$.
$T^{\rm RPA}_{\rm C}(\mu,U_{\rm eff})$ is shown by dashed lines (see text). The arrows mark position of the plateau of the density of states. }
\label{fig:phase_diag_mu_vs_T_t1=-0.13_t2=+0.13}
\end{figure}

We show the obtained phase diagram $T_{\rm C}(\mu, U)$ for the case $\tau = -\tau'=-0.13$ in Fig.~\ref{fig:phase_diag_mu_vs_T_t1=-0.13_t2=+0.13}. The chosen interval of the chemical potentials $\mu$ is taken around vHs plateau, corresponding to the fillings $n\in(0.65,0.68)$ per one spin projection. We also show the line $L_{-}(\mathbf{q} = 0) = 0$ as a function of temperature and $\mu$. 
One can see that as expected, the phase transition temperatures increase with $U$; $T_{\rm C}(\mu,U)$ is also largest for the Fermi level at the narrow van Hove plateau and decrease when moving away from the plateau. The obtained quantum phase transitions are of the first order (corresponding to the sudden disappearance of $T_{\rm C}$ with changing $\mu$) due to change of the sign of $L_{-}({\bf q}=0)$. It is interesting that in~the~considered case the lines of the first-order transitions form themselves the ``quantum critical fan" which begins at $T=0$ at the position of a narrow van Hove plateau. At strong $U$ substantial values of $T_{\rm C}(\mu,U)$ are obtained therefore far beyond the position of the plateau. 


\begin{figure}[b]
\includegraphics[angle=-90,width=.6\textwidth]{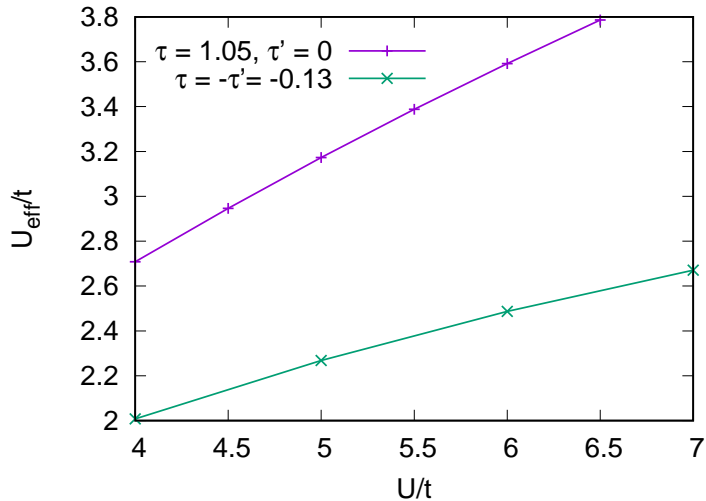}
\caption{$U$ dependence of $U_{\rm eff}$ for the cases $\tau = \tau' = -0.13$ and $\tau = 1.05$, $\tau' = 0$. 
}
\label{fig:U_eff}
\end{figure}

\begin{figure}[t]
\includegraphics[angle=-90,width=.65\textwidth]{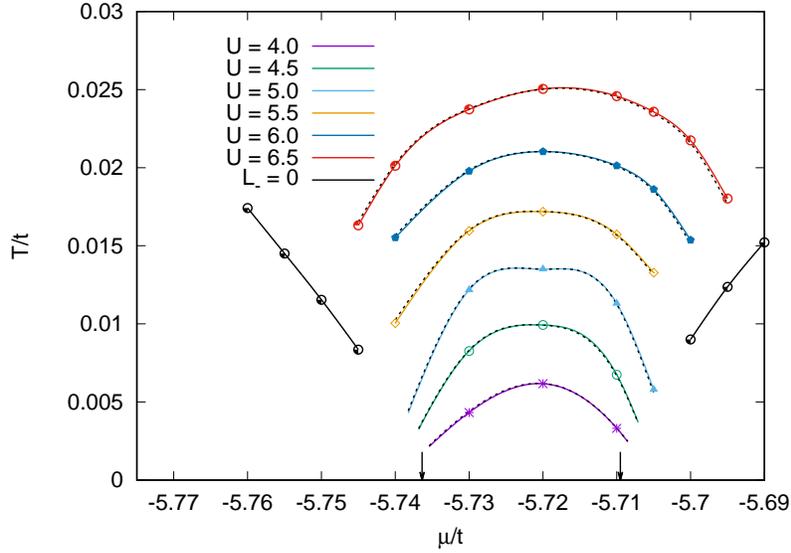}
\caption{The same as Fig.~\ref{fig:phase_diag_mu_vs_T_t1=-0.13_t2=+0.13} at $\tau = 1.05$ and $\tau' = 0$. The value $U_{\rm eff}$ for RPA is adjusted to reproduce the same $T_{\rm C}$ at $\mu = -5.72t$.  The arrows mark position of the plateau of the density of states. 
}
\label{fig:phase_diag_mu_vs_T_t1=1.05_t2=0}
\end{figure}


The obtained values of Curie temperature are almost an order of magnitude smaller than the Curie temperature in RPA. To compare the obtained results with RPA we therefore introduce an effective Coulomb interaction $U_{\rm eff}$ (cf. Ref. \onlinecite{Yuki}). The interaction $U_{\rm eff}$ is determined  to match $T^{\rm RPA}_{\rm C}(\mu_*, U_{\rm eff}) = T_{\rm C}(\mu_*,U)$ for some fixed $\mu=\mu_*$ (for $\tau = -\tau' = -0.13$ we choose $\mu_* = E_{L_2}$). 
From Fig. \ref{fig:phase_diag_mu_vs_T_t1=-0.13_t2=+0.13} one can see that $T^{\rm RPA}_{\rm C}(\mu, U_{\rm eff})$ only weakly deviates from the results of the fRG approach. 
The $U$ dependence of the resulting $U_{\rm eff}$ is shown in~Fig.~\ref{fig:U_eff}. 
We have verified that simple estimate of $U_{\rm eff} = U/(1+U\Pi_{\rm pp}(\mathbf{q} = 0))$ {with $\Pi_{\rm pp}(\mathbf{q}) = \sum_{\mathbf{p}}\{{f[-\xi_{\mathbf{p} - \mathbf{q}/2}] - f[\xi_{\mathbf{p} + \mathbf{q}/2}]}\}/({\xi_{\mathbf{p} + \mathbf{q}/2} + \xi_{\mathbf{p} - \mathbf{q}/2}})$}, corresponding to the summation of ladder  particle-particle diagrams, yields much smaller (up to 4 times) value $U_{\rm eff}$, than obtained within the fRG approach. Despite the fact that this approach was often used for screening strength estimation,  our results imply impossibility of separate renormalization of Coulomb interaction in different channels, which also shows important effect of the interference of different electron scattering channels.

In Fig. \ref{fig:phase_diag_mu_vs_T_t1=1.05_t2=0} we show that the similar dome-shaped form of the phase diagram in terms of $\mu$ and $T_{\rm C}$ is obtained for $\tau = 1.05$ and $\tau' = 0$. 
The chosen interval of chemical potentials $\mu$ is taken around vHs plateau, corresponding to fillings $n\in(0.031,0.056)$ per one spin projection. In this case despite, the relatively broad van Hove plateau, the deviation of $\mu$ from van Hove structure (see Fig.~\ref{fig:dos}(b)) results in a rapid decay of $T_{\rm C}$. 
The role of incommensurate correlations here is the same as in above considered cases for $\tau = -0.13$, they are not enhanced considerably. Choosing $\mu_* = -5.72t$ for the determination of $U_{\rm eff}$, we see from Fig.~\ref{fig:phase_diag_mu_vs_T_t1=1.05_t2=0} that the obtained $T^{\rm RPA}_{\rm C}(\mu, U_{\rm eff})$ for  the case $\tau = 1.05$, $\tau' = 0$ almost perfectly reproduces the results of fRG approach in the region $T_{\rm C}>0$. For the case $\tau = 1.05$, $\tau' = 0$ the renormalization of the interaction $U$ is weaker than for the case $\tau = -\tau' = -0.13$ (see~Fig.~\ref{fig:U_eff}), which seems to be related to smaller density of states in the former case.

\section{Conclusions}\label{sec:conclusions} 
In this study, we present the fRG treatment of an instability towards ferromagnetic order in the Hubbard model, controlled by the Fermi level being in the vicinity of van Hove singularity and the on--site Coulomb interaction.

For the densities of states with the plateau, formed by the higher-order van Hove singularities, we find ferromagnetic instability for the Fermi level being in the vicinity of the plateau, with the Curie temperature decreasing with moving the Fermi level away from the plateau. We find the first-order quantum phase transition to paramagnetic state with further moving the Fermi level away from the plateau.  

In the considered case of van Hove singularities in fcc lattice, incommensurate correlations are not remarkably enhanced. This considerably distinguishes from the case of ferromagnetism formation in two-dimensional case with large $t'/t\sim0.4-0.5$ where strong competition with long-wave incommensurate magnetic fluctuations was found~\cite{2007:Igoshev,2009:Igoshev,2011:Igoshev}. 
We therefore find that in the three-dimensional case the main cause of ferromagnetism suppression is the particle-particle screening. 

The analysis of fRG flows in the the case of logarithmic singularity of density of states ($\tau = -0.13$ and $\tau' = 0.0925$) with the Fermi level being at van Hove singularity yields the absence of ferromagnetic instability well below available temperature scales ($s<s_{\rm max} = 5.5-6.0$). 
One can state that wide plateau with large value of density of states caused by van Hove structure (or, at least, stronger divergence of the density of states) is necessary to stabilize the ferromagnetic instability in the three-dimensional case, and logarithmic van Hove singularity is not sufficient due to strong  particle-particle screening. 

In the present paper we have considered $s$-wave basis function only, which implies averaging of the dependence of vertices on fermionic momenta. In future studies other basis functions, e.g. peaked at the van Hove singularity points, can be included. This may allow to describe the effect of stronger fermionic momentum dependence of vertices near van Hove points.

In realistic materials, the Hund exchange interaction plays important role. On one hand, it provides (partial) formation of local magnetic moments, such that the local dynamic two-particle correlations, not accounted in the present study become important (see, e.g., Refs. \onlinecite{2017:Hausoel,OurFe}). On the other hand, Hund exchange enhances the tendency towards ferromagnetic order, see, in particular, Ref. \onlinecite{Arita}. Studying the effect of Hund exchange in realistic materials within the non-local extensions of dynamical mean-field theory \cite{OurRev}, in particular, the DMF$^2$RG approach \cite{DMF2RG}, is therefore of certain interest.


\section{Acknowledgments}\label{sec:acknowledgments} 
The research was carried out within the state assignment of the Ministry of Science and Higher Education of the Russian Federation (theme  ``Quant'' No. 122021000038-7) and partly supported by RFBR grant~20-02-00252a. A.~K.~also acknowledges the support from the
Ministry of Science and Higher Education of the Russian Federation (Agreement No.~075-15-2021-606). 
The calculations were performed on the Uran supercomputer at the IMM UB RAS. 
\appendix
\section{Electron dispersion and giant van Hove singularities of DOS}\label{sec:spectrum_and_DOS}

\begin{figure}[tbp]
\includegraphics[angle=-90,width=.6\textwidth]{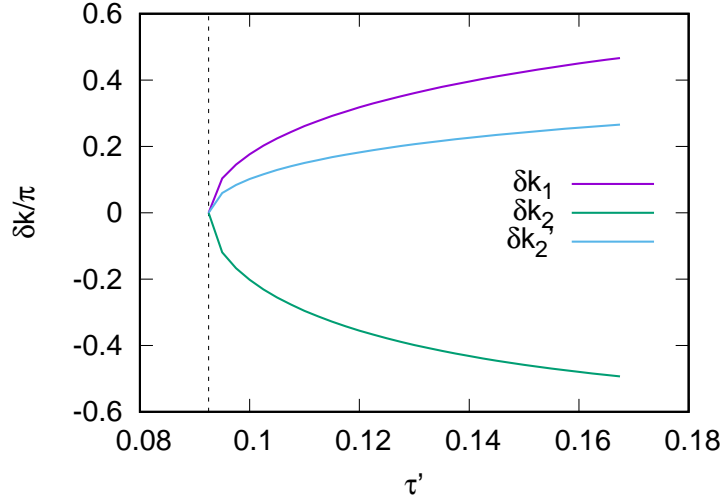}
\caption{$k$ deviations of van Hove position, see Eqs.~(\ref{eq:delta_k1})--(\ref{eq:delta_k2'}). 
}
\label{fig:vHS_appendix}
\end{figure}

At $\tau' < \tau'_{\rm c}$, only van Hove L point is present. 
The investigation of the dispersion (\ref{eq:bare_spectrum}) in the vicinity of L~point at $\tau'_{\rm c} < \tau'$ yields the following van Hove points: (i) L$_1(\pi - \delta k_1, \pi + \delta k_1, \pi)$ (on the face of Brillouin zone), where 
\begin{equation}\label{eq:delta_k1}
\delta k_1(\tau, \tau') = 2\arcsin\sqrt{1 - \frac{\tau'_{\rm c}(\tau)}{\tau'}}.
\end{equation} 
(ii) L$_2(\pi - \delta k_2(\tau, \tau'), \pi - \delta k_2'(\tau, \tau'), \pi - \delta k_2'(\tau, \tau'))$, where
\begin{eqnarray}
\label{eq:delta_k2}
\delta k_2(\tau, \tau')  &=& -2\arcsin \frac{2\mathfrak{s}(\tau, \tau')}{1 + 8\delta\tau'(\tau) - 16\tau'\mathfrak{s}^2(\tau, \tau')}, \\
\label{eq:delta_k2'}
\delta k_2'(\tau, \tau') &=& 2\arcsin \mathfrak{s}(\tau, \tau'),
\end{eqnarray}
where $\delta\tau'(\tau) = \tau' - \tau'_{\rm c}(\tau)$ 
and $\mathfrak{s} = \mathfrak{s}(\tau, \tau')$ is a root of the equation
\begin{equation}
\delta\tau' + \tau'\mathfrak{s}^2 + 4\left(1 +  4\delta\tau'(\tau) - 4\tau'\mathfrak{s}^2\right)(-\delta\tau'(\tau) + 2\tau'\mathfrak{s}^2)(1 + 4\delta\tau'(\tau) - 8\tau'\mathfrak{s}^2 ) = 0.
\end{equation}
At small $\delta\tau'$, we can simplify this equation by neglecting small terms  in brackets compared to~1, to obtain
\begin{equation}
-3\delta\tau'(\tau) + 9\tau'\mathfrak{s}^2 = 0,
\end{equation}
\begin{eqnarray}
\delta k_2(\tau, \tau')  &\approx& -2\arcsin 2\sqrt{\frac{\delta\tau'}{3\tau'}}, \\
\delta k_2'(\tau, \tau') &\approx& 2\arcsin\sqrt{\frac{\delta\tau'}{3\tau'}}.
\end{eqnarray}
Note that at small $\delta\tau'$ we have $\delta k_2(\tau, \tau')  \approx 2\delta k_2'(\tau, \tau') $ which implies that L$_2$ point stays closely to the face of Brillouin zone at small $\delta\tau'$.  
The energies of these van Hove points as a function of $\tau'$ are shown in~Fig.~\ref{fig:vHS} in the main text. The deviations $\delta k_1(\tau, \tau')$, $\delta k_2(\tau, \tau')$, $\delta k_2'(\tau, \tau')$ as functions of $\tau$ are shown in Fig.~\ref{fig:vHS_appendix}. 
One can see that the energy levels of $\k_{\rm L_1}$, $\k_{\rm L_2}$ points are very close and the distance between these points increases slowly with increasing of $\tau'$. This scenario of high-order van Hove point destruction was earlier found for cubic lattices for the dispersion in~the~nearest and next-nearest approximation~\cite{2019:Igoshev_PMM,2019:Igoshev_JETP,2022:Igoshev}.

In the case $\tau' = \tau'_{\rm c}$, see Eq.~(\ref{eq:giant_vHS_condition}), we expand the dispersion in the vicinity of L point up to fourth-order terms with respect to both $k_\Lambda$ and $q_\perp$   
\begin{equation}
    \epsilon_{\mathbf{k}_{\rm L}+\Delta\mathbf{k}} = \frac32(1-2\tau) - \frac32k^2_\varLambda + \frac{1}{96}\left(
    12 k^2_\varLambda q^2_\perp + 8 (1 + \tau) k^4_\varLambda + 3 (1 + 2 \tau) q^4_\perp + 
 2 \sqrt{2} (1 + 4 \tau) \sin3\varphi k_\varLambda q^3_\perp
    \right),
\end{equation}
here we parameterize the deviation from L point as $\Delta\mathbf{k} = q_\perp\cos\varphi\mathbf{e}_1 + q_\perp\sin\varphi\mathbf{e}_2 + k_\varLambda\mathbf{e}_3$, where the orthonormal basis $\mathbf{e}_1(+1/\sqrt{2},0,-1/\sqrt{2})$, $\mathbf{e}_2(-1/\sqrt{6},+2/\sqrt{6},-1/\sqrt{6})$, $\mathbf{e}_3(+1/\sqrt{3},+1/\sqrt{3},+1/\sqrt{3})$ was used. 
For convenience, we retain only leading contributions into $\epsilon_{\mathbf{k}_{\rm L}+\Delta\mathbf{k}}$ with respect to $k_\Lambda$ and $q_\perp$ dependence:
\begin{equation}\label{eq:simplified_expansion}
\delta\epsilon_{\mathbf{k}_{\rm L}+\Delta\mathbf{k}} \approx \epsilon_{\mathbf{k}_{\rm L}+\Delta\mathbf{k}} - \frac32(1-2\tau) =  -\frac32k^2_\varLambda + \frac{1+2\tau}{32}q^4_\perp.        
\end{equation}
It can be verified that neglected terms in Eq.~(\ref{eq:simplified_expansion}) do not change substantially the $\delta\epsilon_{\mathbf{k}_{\rm L}+\Delta\mathbf{k}}$ in a rather extended vicinity of the L point. 
Since here $q_\perp^2$ plays a similar role as $k_\Lambda$ we introduce the same (up to numerical factors) cutoff $\kappa$ for them: 
\begin{equation}
(3/2)k_\Lambda^2, \frac{1+2\tau}{32}q^4_\perp < \kappa^2 \lesssim 1,
\end{equation}
From the DOS definition we get the contribution of 8 regions near the L point and corresponding points shifted by reciprocal lattice constants. 
We obtain:
\begin{equation}
\Delta \rho(\delta\epsilon) \approx \frac{8\pi}{V_{\rm BZ}}\int\limits_{-\sqrt{2/3}\kappa}^{\sqrt{2/3}\kappa}dk_\Lambda\int\limits_0^{\sqrt{32\kappa/(1+2\tau)}}dq_\perp q_\perp\delta\left(\delta\epsilon +\frac32k^2_\varLambda - \frac{1+2\tau}{32}q^4_\perp\right),
\end{equation}
where $V_{\rm BZ} = 2\pi^3$ being Brillouin zone volume. 
So we obtain that in this case the logarithmic giant van Hove singularity 
\begin{equation}
\Delta \rho(\delta\epsilon) \approx \frac{8}{\pi^2\sqrt3\sqrt{1+2\tau}}\ln\frac{4\kappa^2}{|\delta\epsilon|}.
\end{equation}

\section{RG equations\label{AppB}}
We consider the one-loop truncation of fRG equations for $V(p_1,p_2,p_3)$~\cite{fRG}.

\begin{equation}\label{eq:main_fRG}
    \partial_s V(p_1, p_2, p_3) = \tau_{\rm pp}(p_1, p_2, p_3) + \tau^{\rm d}_{\rm ph}(p_1, p_2, p_3) + \tau^{\rm cr}_{\rm ph}(p_1, p_2, p_3),
\end{equation}
where
\begin{align}
    \label{eq:tpp}
    \tau_{\rm pp}(p_1,p_2,p_3) &= -\sum_p \partial_s\left[G(p)G(p_1+p_2 - p)\right]V\left(p_1,p_2,p\right)V\left(p_1+p_2-p,p,p_3\right),\\
    \label{eq:tphcr}
    \tau_{\rm ph}^{\rm cr}(p_1,p_2,p_3) &= -\sum_p \partial_s\left[G(p)G(p+p_3 - p_1)\right]V\left(p_1,p+p_3-p_1,p_3\right)V\left(p,p_2,p+p_3-p_1\right)
\end{align}
\begin{equation}
\begin{aligned}
        \label{eq:tphd}
        \tau_{\rm ph}^{\rm d}(p_1,p_2,p_3)  & = \sum_p\partial_s\left[G(p)G(p+p_2 - p_3)\right] 
        [2V\left(p_1,p+p_2-p_3,p\right)V\left(p,p_2,p_3\right) \\ 
        & - V\left(p_1,p+p_2-p_3,p_1+p_2-p_3\right)V\left(p,p_2,p_3\right) \\ 
        & -V\left(p_1,p+p_2-p_3,p\right)V\left(p,p_2,p+p_2-p_3\right)].
\end{aligned}
\end{equation}
Here $s$ is fRG scale parameter, $p = (\I\nu_n, \mathbf{p})$ is a 3-momentum, $\nu_n = \pi(2n + 1)T$ being the fermionic Matsubara frequency with integer $n_p$ and $\mathbf{p}$ being a vector in the Brillouin zone. Analogous notations hold for $p_i$. 
We use the shorthand notation 
\begin{equation}
\sum_p = {T}\sum_{\nu_n,\mathbf{p}}.
\end{equation}
Three contribution in Eq.~(\ref{eq:main_fRG}) corresponds to particle-particle $\tau_{\rm pp}$~(see Eq.~(\ref{eq:tpp})), direct particle-hole $\tau_{\rm ph}^{\rm d}$~(see Eq.~(\ref{eq:tphd})) and crossed particle-hole direct particle-hole $\tau_{\rm ph}^{\rm cr}$~(see Eq.~(\ref{eq:tphcr})). 

We rewrite the~Eq.~(\ref{eq:tpp}) through momenta $l,k_1,k_2$, introduced for particle-particle (P) channel by~Eq.~(\ref{eq:SC_momentum_def}), performing the variable change~
$p\rightarrow l/2 - p$
\begin{align}
    \tau_{\rm pp}(l/2 + k_1,l/2 - k_1, l/2 - k_2) &= \sum_p L_{\rm pp}(l,p)V\left(l/2 + k_1,l/2 - k_1, l/2 - p\right)\notag\\
    &\times V\left(l/2 + p,l/2 - p, l/2 - k_2\right).
\end{align}
Here arguments of both $V$ factors are the same as the arguments on the left hand side. Following Ref.~\onlinecite{2017:TUfRG}, we use the identity
\begin{equation}\label{eq:delta_identity}
    \delta_{\mathbf{k}\mathbf{k}'} = \sum_nf_n(\mathbf{k})f_n(\mathbf{k}')
\end{equation}
to obtain \begin{align}
    \tau_{\rm pp}(l/2 + k_1,l/2 &- k_1, l/2 - k_2) = \sum_{m_1,m_2,n_1,n_2}f_{n_1}(\mathbf{k}_1)f_{n_2}(\mathbf{k}_2)
    \sum_p f_{m_1}(\mathbf{p})L_{\rm pp}(l,p)f_{m_2}(\mathbf{p})\notag\\
    &\times\sum_{\mathbf{k}_1'\mathbf{p}_1'} f_{n_1}(\mathbf{k}_1')V\left(l/2 + \mathbf{k}_1',l/2 - \mathbf{k}_1', l/2 - \mathbf{p}_1'\right)f_{m_1}(\mathbf{p}_1')\notag\\
    &\times\sum_{\mathbf{k}_2'\mathbf{p}_2'}
    f_{n_2}(\mathbf{k}_2')V\left(l/2 + \mathbf{p}_2',l/2 - \mathbf{p}_2', l/2 - \mathbf{k}_2'\right)f_{m_2}(\mathbf{p}_2').
\end{align}
Following the same algorithm for the crossed particle-hole (M) channel, 
for the contribution (\ref{eq:tphcr}), we get
\begin{align}
    \tau_{\rm ph}^{\rm cr}(l/2 + k_1,k_2 &- l/2, k_1 - l/2) = \sum_{m_1,m_2,n_1,n_2}f_{n_1}(\mathbf{k}_1)f_{n_2}(\mathbf{k}_2)
    \sum_p f_{m_1}(\mathbf{p})L_{\rm ph}(l,p)f_{m_2}(\mathbf{p})\notag\\
    &\times\sum_{\mathbf{k}_1'\mathbf{p}_1'} f_{n_1}(\mathbf{k}_1')V\left(l/2 + \mathbf{k}_1',\mathbf{p}_1' - l/2, \mathbf{k}_1' - l/2\right)f_{m_1}(\mathbf{p}_1')\notag\\
    &\times\sum_{\mathbf{k}_2'\mathbf{p}_2'}
    f_{n_2}(\mathbf{k}_2')V\left(l/2 + \mathbf{p}_2',\mathbf{k}_2' - l/2, \mathbf{p}_2' - l/2 \right)f_{m_2}(\mathbf{p}_2').
\end{align}
Now we consider the projection of the expression
(\ref{eq:tphd}) onto the channel~K (see~Eq.~(\ref{eq:K_momentum_def})). 
We have
\begin{align}
    \tau_{\rm ph}^{\rm d}(l/2 + k_1,k_2 &- l/2, l/2 + k_2) = -\sum_{m_1,m_2,n_1,n_2}f_{n_1}(\mathbf{k}_1)f_{n_2}(\mathbf{k}_2)
    \sum_p f_{m_1}(\mathbf{p})L_{\rm ph}(l,p)f_{m_2}(\mathbf{p})\notag\\
    &\times\sum_{\mathbf{k}_1'\mathbf{p}_1'\mathbf{k}_2'\mathbf{p}_2'}\left[2f_{n_1}(\mathbf{k}_1')V\left(l/2 + \mathbf{k}_1',\mathbf{p}_1' - l/2, \mathbf{p}_1' + l/2\right)f_{m_1}(\mathbf{p}_1')
    \right. \notag\\ &\left.
    \times
    f_{n_2}(\mathbf{k}_2')V\left(\mathbf{p}_2' + l/2,\mathbf{k}_2' - l/2, l/2 + \mathbf{k}_2' \right)f_{m_2}(\mathbf{p}_2')
    \right. \notag\\ &\left.
    - f_{n_1}(\mathbf{k}_1')V\left(l/2 + \mathbf{k}_1',\mathbf{p}_1' - l/2, \mathbf{k}_1' - l/2\right)f_{m_1}(\mathbf{p}_1')
    \right. \notag\\ &\left.
    \times
    f_{n_2}(\mathbf{k}_2')V\left(\mathbf{p}_2' + l/2,\mathbf{k}_2' - l/2, l/2 + \mathbf{k}_2' \right)f_{m_2}(\mathbf{p}_2')
    \right. \notag\\ &\left.
    - f_{n_1}(\mathbf{k}_1')V\left(l/2 + \mathbf{k}_1',\mathbf{p}_1' - l/2, \mathbf{p}_1' + l/2\right)f_{m_1}(\mathbf{p}_1')
    \right. \notag\\ &\left.
    \times
    f_{n_2}(\mathbf{k}_2')V\left(\mathbf{p}_2' + l/2,\mathbf{k}_2' - l/2, \mathbf{p}_2' - l/2 \right)f_{m_2}(\mathbf{p}_2')\right].
\end{align}
The second term in Eq.~(\ref{eq:main_K}) contains momentum arguments suitable for projection in K channel 
\begin{multline}
    \tau^{\rm cr}_{\rm ph}(k_1 + l/2, k_2 - l/2, k_1 - l/2) =  \sum_p L_{\rm ph}(l,p)V\left(k_1 + l/2, p - l/2, k_1 - l/2\right)\\
    \times V\left(p + l/2, k_2 - l/2, p - l/2 \right).
\end{multline}
The third momentum argument of both $V$ factors does not have the form of K-channel, see Eq.~(\ref{eq:K_projection}), therefore we,  as~above, interchange third and forth momentum, $V\left(k_1 + l/2, p - l/2, k_1 - l/2\right) = V_{\rm ex}\left(k_1 + l/2, p - l/2, p + l/2\right)$, 
$V\left(l/2 + p,k_2 - l/2, p - l/2 \right) = V_{\rm ex}\left(l/2 + p,k_2 - l/2, k_2 + l/2 \right)$.
Again using the identity~(\ref{eq:delta_identity}), we get
\begin{align}
    \tau^{\rm cr}_{\rm ph}(l/2 + k_1,k_2 &- l/2, k_1 - l/2) = 
    \sum_{n_1,n_2}f_{n_1}(\mathbf{k}_1)f_{n_2}(\mathbf{k}_2)
    \sum_p f_{m_1}(\mathbf{p})f_{m_2}(\mathbf{p}) L_{\rm ph}(l,p)\notag\\
    &\times \sum_{\mathbf{k}_1'\mathbf{p}_1'} f_{n_1}(\mathbf{k}_1')f_{m_1}(\mathbf{p}_1')V_{\rm ex}\left(l/2 + \mathbf{k}_1',\mathbf{p}_1' - l/2, \mathbf{p}_1' + l/2\right)\notag\\
    &\times
    \sum_{\mathbf{k}_2'\mathbf{p}_2'} f_{n_2}(\mathbf{p}_2')f_{m_2}(\mathbf{k}_2')V_{\rm ex}\left(\mathbf{p}_2' + l/2, \mathbf{k}_2' - l/2, \mathbf{k}_2' + l/2\right).
\end{align}

We consider the projection of the interaction in Eq.~(\ref{eq:V}). {The dependence of vertex $V$ on fermionic frequencies of $k_1$ and $k_2$ in all channels is out of the scope within this projection procedure and is not considered in our study.}  
For convenience we introduce 
\begin{equation}
\Phi_{\rm M-K} = \Phi_{\rm M} - \Phi_{\rm K}, 
\end{equation}
using the fact that the contributions of channels M and K enters the last two terms {of Eq.~(\ref{eq:V})} in the same way. 
For P-channnel projection we use the representation
\begin{multline}
        V_{\rm P}(\mathbf{l}; \mathbf{k}_1, \mathbf{k}_2) = U - \Phi_{\rm SC}\left(\mathbf{l}, \mathbf{k}_1, \mathbf{k}_2\right) 
        + \Phi_{\rm M}\left(\mathbf{k}_1 + \mathbf{k}_2, \dfrac{\mathbf{l} + \mathbf{k}_1 - \mathbf{k}_2}{2}, \dfrac{\mathbf{l} - \mathbf{k}_1 + \mathbf{k}_2}{2}\right)\\
        + \dfrac{1}{2}\Phi_{\rm M - K}\left(\mathbf{k}_1 - \mathbf{k}_2, \dfrac{\mathbf{l} + \mathbf{k}_1 + \mathbf{k}_2}{2}, \dfrac{\mathbf{l} - \mathbf{k}_1 - \mathbf{k}_2}{2}\right),
\end{multline}
therefore we derive Eq.~(\ref{eq:Phi_SC_final}) with
\begin{multline}
    V^{\rm P}_{n_1n_2}(\mathbf{l}) = U\delta_{n_1,0}\delta_{n_2,0} -  \Phi^{\rm SC}_{n_1n_2}(\mathbf{l}) 
    + \sum_{\k_1\k_2} f_{n_1}(\mathbf{k}_1)f_{n_2}(\mathbf{k}_2)\Phi_{\rm M}\left(\mathbf{k}_1 + \mathbf{k}_2, \dfrac{\mathbf{l} + \mathbf{k}_1 - \mathbf{k}_2}{2}, \dfrac{\mathbf{l} - \mathbf{k}_1 + \mathbf{k}_2}{2}\right)\\
    + \dfrac{1}{2}\sum_{\k_1\k_2} f_{n_1}(\mathbf{k}_1)f_{n_2}(\mathbf{k}_2)\Phi_{\rm M - K}\left(\mathbf{k}_1 - \mathbf{k}_2, \dfrac{\mathbf{l} + \mathbf{k}_1 + \mathbf{k}_2}{2}, \dfrac{\mathbf{l} - \mathbf{k}_1 - \mathbf{k}_2}{2}\right). 
\end{multline}
Analogously for M-channel we derive Eq.~(\ref{eq:Phi_M_final}) with
\begin{multline}
        V_{\rm M}(\mathbf{l}; \mathbf{k}_1, \mathbf{k}_2) = U - \Phi_{\rm SC}\left(\mathbf{k}_1 + \mathbf{k}_2, \dfrac{\mathbf{l} + \mathbf{k}_1 - \mathbf{k}_2}{2}, \dfrac{\mathbf{l} + \mathbf{k}_2 - \mathbf{k}_1}{2}\right) 
        + \Phi^{\rm M}\left(\mathbf{l}, \mathbf{k}_1, \mathbf{k}_2\right)\\
        + \dfrac{1}{2}\Phi_{\rm M - K}\left(\mathbf{l} + \mathbf{k}_1 - \mathbf{k}_2, \dfrac{\mathbf{l} + \mathbf{k}_1 + \mathbf{k}_2}{2}, \dfrac{-\mathbf{l} + \mathbf{k}_1 + \mathbf{k}_2}{2}\right),
\end{multline}
and 
\begin{multline}
        V^{\rm M}_{n_1n_2}(\mathbf{l}) = U\delta_{n_1,0}\delta_{n_2,0}  - \sum_{\k_1\k_2} f_{n_1}(\mathbf{k}_1)f_{n_2}(\mathbf{k}_2)\Phi_{\rm SC}\left(\mathbf{k}_1 + \mathbf{k}_2, \dfrac{\mathbf{l} + \mathbf{k}_1 - \mathbf{k}_2}{2}, \dfrac{\mathbf{l} + \mathbf{k}_2 - \mathbf{k}_1}{2}\right) + \Phi^{\rm M}_{n_1n_2}(\mathbf{l}) \\
        + \dfrac{1}{2}\sum_{\k_1\k_2} f_{n_1}(\mathbf{k}_1)f_{n_2}(\mathbf{k}_2)\Phi_{\rm M - K}\left(\mathbf{l} + \mathbf{k}_1 - \mathbf{k}_2, \dfrac{\mathbf{l} + \mathbf{k}_1 + \mathbf{k}_2}{2}, \dfrac{-\mathbf{l} + \mathbf{k}_1 + \mathbf{k}_2}{2}\right).
\end{multline}
Analogously for K-channel we derive Eq.~(\ref{eq:Phi_K_final}) with
\begin{multline}
        V_{\rm K}(\mathbf{l}; \mathbf{k}_1, \mathbf{k}_2) = U -\Phi_{\rm SC}\left(\mathbf{k}_1 + \mathbf{k}_2, \dfrac{\mathbf{l} + \mathbf{k}_1-\mathbf{k}_2}{2}, \dfrac{-\mathbf{l} + \mathbf{k}_1 - \mathbf{k}_2}{2}\right) \\
        + \Phi_{\rm M}\left(\mathbf{k}_1 - \mathbf{k}_2, \dfrac{\mathbf{l} + \mathbf{k}_1 + \mathbf{k}_2}{2}, \dfrac{-\mathbf{l} + \mathbf{k}_1 + \mathbf{k}_2}{2}\right) 
        + \dfrac{1}{2}\Phi_{\rm M - K}\left(\mathbf{l}, \mathbf{k}_1, \mathbf{k}_2\right)
\end{multline}
and 
\begin{multline}
    V^{\rm K}_{n_1n_2}(\mathbf{l}) = U\delta_{n_1,0}\delta_{n_2,0} - \sum_{\k_1\k_2} f_{n_1}(\mathbf{k}_1)f_{n_2}(\mathbf{\mathbf{k}}_2)\Phi_{\rm SC}\left(\mathbf{k}_1 + \mathbf{k}_2, \dfrac{\mathbf{l} + \mathbf{k}_1-\mathbf{k}_2}{2}, \dfrac{-\mathbf{l} + \mathbf{k}_1 - \mathbf{k}_2}{2}\right) \\
    + \sum_{\k_1\k_2} f_{n_1}(\mathbf{k}_1)f_{n_2}(\mathbf{k}_2)\Phi_{\rm M}\left(\mathbf{k}_1 - \mathbf{k}_2, \dfrac{\mathbf{l} + \mathbf{k}_1 + \mathbf{k}_2}{2}, \dfrac{-\mathbf{l} + \mathbf{k}_1 + \mathbf{k}_2}{2}\right)
    + \dfrac{1}{2}\Phi^{\rm M - K}_{n_1n_2}(\mathbf{l}).
\end{multline}

\section{Single-boson exchange decomposition\label{AppC}}

Eqs.~(\ref{eq:dot_Phi_SC})-(\ref{eq:dot_Phi_K}) allow single-boson exchange decomposition of the vertex function through triangular (Hedin, ``Yukawa'') vertex functions  $\gamma_{\rm SC,M,K}(\q)$ and bosonic propagators $D_{\rm SC,M,K}(\q)$~\cite{2022:BFfRG}
\begin{eqnarray}
U - \Phi_{\rm SC}(\q) &=& \gamma^2_{\rm SC}(\q)D_{\rm SC}(\q) + R_{\rm SC}(\q), \\
U + \Phi_{\rm M}(\q) &=& \gamma^2_{\rm M}(\q)D_{\rm M}(\q) + R_{\rm M}(\q), \\
U - \Phi_{\rm K}(\q) &=& \gamma^2_{\rm K}(\q)D_{\rm K}(\q) + R_{\rm K}(\q).
\end{eqnarray}
where the triangular vertex functions assumed to be independent on fermionic momenta.   
Retaining of residue contributions $R_{\rm SC,M,K}(\q)$ correspond to taking into account of so-called ``box'' diagrams for 4-vertex (their dependence on fermionic momenta is also neglected). 

Instead of one equation in each channel we get 3 ones (for $D$, $\gamma$ and $R$). The equations for $D$ read
\begin{eqnarray}
	\dot D_{\rm SC}(\q) &=& \gamma^2_{\rm SC}(\q)D_{\rm SC}^2(\q) L_{+}(\q),\\
	\dot D_{\rm M}(\q) &=& \gamma^2_{\rm M}(\q)D_{\rm M}^2(\q) L_{-}(\q),\\
	\dot D_{\rm K}(\q) &=& -\gamma^2_{\rm K}(\q)D_{\rm K}^2(\q)L_{-}(\q).
\end{eqnarray}
The equation for $\gamma$ read
\begin{eqnarray}
	\dot \gamma_{\rm SC}(\q) &=& \gamma_{\rm SC}(\q)
	\left(R_{\rm SC}(\q) + \frac32\bar\Phi_{\rm M} - \frac12\bar\Phi_{\rm K}\right) L_{+}(\q),\\
	\dot\gamma_{\rm M}(\q) &=& \gamma_{\rm M}(\q)
	\left(R_{\rm M}(\q) - \bar{\Phi}_{\rm SC} + \frac12\bar{\Phi}_{\rm M} - \frac12\bar{\Phi}_{\rm K}\right) L_{-}(\q),\\
	\dot \gamma_{\rm K}(\q) &=& -\gamma_{\rm K}(\q)
	\left(R_{\rm K}(\q) - \bar{\Phi}_{\rm SC} + \frac32\bar{\Phi}_{\rm M} + \frac12\bar{\Phi}_{\rm K}\right)L_{-}(\q).
\end{eqnarray}
The equation for $R$ read
\begin{eqnarray}
	\dot R_{\rm SC}(\q) &=& 
	\left(R_{\rm SC}(\q) + \frac32\bar\Phi_{\rm M} - \frac12\bar\Phi_{\rm K}\right)^2 L_{+}(\q),\\
	\dot R_{\rm M}(\q) &=& 
	\left(R_{\rm M}(\q) - \bar{\Phi}_{\rm SC} + \frac12\bar{\Phi}_{\rm M} - \frac12\bar{\Phi}_{\rm K}\right)^2 L_{-}(\q),\\
	\dot R_{\rm K}(\q) &=& 
	-\left(R_{\rm K}(\q) - \bar{\Phi}_{\rm SC} + \frac32\bar{\Phi}_{\rm M} + \frac12\bar{\Phi}_{\rm K}\right)^2L_{-}(\q).
\end{eqnarray}


\begin{thebibliography}{99}

\bibitem{1958:ZrZn2_discovery} B. T. Matthias and R. M. Bozorth, Phys. Rev. \textbf{109}, 604 (1958).

\bibitem{2020:Skornyakov} S. L. Skornyakov, V. S. Protsenko, V. I. Anisimov, and A. A. Katanin, Phys. Rev. B~\textbf{102}, 085101~(2020).


\bibitem{1984:Ni3Al:Sasakura} H. Sasakura, K. Suzuki, Y. Masuda, Journal of the Physical Society of Japan~\textbf{53}, 352~(1984).

\bibitem{2005:Niklowitz} P. G. Niklowitz, F. Beckers, G. G. Lonzarich, G. Knebel, B. Salce, J. Thomasson, N. Bernhoeft, D. Braithwaite, and J. Flouquet, Phys. Rev. B~\textbf{72}, 024424~(2005).

\bibitem{Stoner} E. C. Stoner Proc. Roy. Soc. A \textbf{165} (922), 372 (1938); \textbf{169} (938), 339 (1939).

\bibitem{1966:Penn} D. R. Penn, Phys. Rev. \textbf{142}, 350 (1966).

\bibitem{1972:Murata} 
K. K. Murata and S. Doniach, Phys. Rev. Lett.~\textbf{29}, 285~(1972).

\bibitem{Moriya_book} T. Moriya, \textit{Spin Fluctuations in Itinerant Electron Magnetism}, Springer: Berlin, 1985.

\bibitem{1976:Hertz} J. A. Hertz, Phys. Rev. B \textbf{14}, 1165~(1976).

\bibitem{1993:Millis} A. J. Millis, Phys. Rev.~B~\textbf{48}, 7183~(1993).

\bibitem{2001:Katanin} V. Yu. Irkhin, A. A. Katanin, and M. I. Katsnelson, Phys. Rev. B 64, 165107 (2001).

\bibitem{2001:Salmhofer} C. Honerkamp and M. Salmhofer, Phys. Rev. Lett. {\bf 87}, 187004 (2001); Phys. Rev. B {\bf 64}, 184516 (2001).

\bibitem{Kampf} A. A. Katanin and A. P. Kampf, Phys. Rev. B \textbf{68}, 195101 (2003).


\bibitem{1993:Katsnelson}
S. V. Vonsovskii, M. I. Katsnelson, and A. V. Trefilov, Fiz. Met. Metalloved. 76 (3) 3 (1993); 76 (4), 3 (1993).

\bibitem{2019:Igoshev_PMM} P.~A.~Igoshev and V.~Yu.~Irkhin, Phys.  Met. Metallogr.~\textbf{120}, 1282~(2019).

\bibitem{2019:Igoshev_JETP} P.~A.~Igoshev and V.~Yu.~Irkhin, JETP Lett.~\textbf{110}, 727~(2019).

\bibitem{2020:Classen} L. Classen, A. V. Chubukov, C. Honerkamp, M. M. Scherer, Phys. Rev. B 102, 125141 (2020).

\bibitem{2022:Igoshev} P.A. Igoshev, V.Yu. Irkhin, Physics Letters A \textbf{438}, 128107~(2022).

\bibitem{2022:Betouras} A. Chandrasekaran, J. J. Betouras, ArXiv: 2207.06099.


\bibitem{2007:Yamaji}
Y.~Yamaji, T.~Misawa, and M. Imada, Journal of~the~Physical Society of Japan~\textbf{76}, 063702~(2007).

\bibitem{2004:Major} Zs. Major, S.~B.~Dugdale, R.~J.~Watts, G.~Santi, M.~A.~Alam, S.~M.~Hayden, J.~A.~Duffy, J.~W.~Taylor, T.~Jarlborg, E.~Bruno, D.~Benea, and H.~Ebert, Phys. Rev. Lett.~\textbf{92}, 107003 (2004).

\bibitem{2003:Yates} S.~J.~C.~Yates, G.~Santi, S.~M.~Hayden, P.~J.~Meeson, and~S.~B.~Dugdale, Phys. Rev. Lett.~\textbf{90}, 057003~(2003).


\bibitem{2011:Hamid} A. S. Hamid, A. Uedono, Zs. Major, T. D. Haynes, J. Laverock, M. A. Alam, S. B. Dugdale, and D. Fort, Phys. Rev. B~\textbf{84}, 235107 (2011).

\bibitem{2017:Hausoel} A. Hausoel, M. Karolak, E. Sasioglu, A. Lichtenstein, K. Held, A. Katanin, A. Toschi, and G. Sangiovanni,  Nat. Commun. \textbf{8}, 16062 (2017). 

\bibitem{OurFe} A. A. Katanin, A. I. Poteryaev, A. V. Efremov, A. O. Shorikov, S. L. Skornyakov, M. A. Korotin, and V. I. Anisimov, Phys. Rev. B {\bf 81}, 045117 (2010).

\bibitem{2015:Efremov} P. A. Igoshev, A. V. Efremov, A. A. Katanin, 	Phys. Rev. B \textbf{91}, 195123 (2015).

\bibitem{Kanamori} J. Kanamori, Progr. Theor. Phys. {\bf 30}, 275 (1963).

\bibitem{Tmatrix1} R. Hlubina, Phys. Rev. B {\bf 59}, 9600 (1999). 

\bibitem{3body1} C. Calandra and F. Manghi, Phys. Rev. B {\bf 50}, 2061 (1994); F. Manghi, V. Bellini and C. Arcangeli, Phys. Rev. B {\bf 56}, 7149 (1997).

\bibitem{3body2} J. Igarashi, J. Phys. Soc. Jpn. {\bf 52}, 2827 (1983); J. Igarashi, M. Takahashi, and T. Nagano, J. Phys. Soc. Jpn. {\bf 68}, 3682 (1999).  

\bibitem{Ulmke} M. Ulmke, Eur. Phys. J. B {\bf 1}, 301 (1998); D. Vollhardt, N. Blümer, K. Held, M. Kollar, J. Schlipf, M. Ulmke, J. Wahle,  Advances in Solid State Physics {\bf 38} (Vieweg, Wiesbaden, 1999), p. 383.

\bibitem{Pruschke} R. Zitzler, Th. Pruschke, and R. Bulla, Eur. Phys. J. B {\bf 27}, 473-481 (2002).

\bibitem{Licht} A. I. Lichtenstein, M. I. Katsnelson, and G. Kotliar, Phys. Rev. Lett. {\bf 87}, 067205 (2001).

\bibitem{Arita} S. Sakai, R. Arita, and H. Aoki, Phys. Rev. Lett. {\bf 99}, 216402 (2007).

\bibitem{Schulz} H. J. Schulz, Phys. Rev. Lett. \textbf{64}, 1445 (1990).

\bibitem{2010:Igoshev} P. A. Igoshev, M. A. Timirgazin, A. A. Katanin, A. K. Arzhnikov, and V. Yu. Irkhin, Phys. Rev. B \textbf{81}, 094407 (2010)

\bibitem{2015:Igoshev} P.A.~Igoshev, M.A.~Timirgazin, V.F.~Gilmutdinov, A.K.~Arzhnikov and V. Yu.~Irkhin, J. Phys.: Cond. Matt. \textbf{27}, 446002 (2015).

\bibitem{fRG} M. Salmhofer and C. Honerkamp, Prog. Theor. Phys. \textbf{105}, 1 (2001).

\bibitem{Zanchi} D. Zanchi and H. J. Schulz, Phys. Rev. B \textbf{61}, 13609 (2000).

\bibitem{Halboth} C. J. Halboth and W. Metzner, Phys. Rev. B \textbf{61}, 7364 (2000).

\bibitem{Honerkamp01} C. Honerkamp, M. Salmhofer, N. Furukawa, and T. M. Rice, Phys. Rev. B {\bf 63}, 035109 (2001).

\bibitem{Honerkamp} C. Honerkamp, D. Rohe, S. Andergassen, and T. Enss, Phys. Rev. B \textbf{70}, 235115 (2004).

\bibitem{Rohe} W. Metzner, J. Reiss, and D. Rohe, Phys. Stat. Sol.(b) \textbf{243}, 46 (2005).

\bibitem{Lauscher} M. Salmhofer, C. Honerkamp, W. Metzner, and O. Lauscher, Prog. Theor. Phys. \textbf{112}, 943 (2004); R. Gersch, C. Honerkamp, D. Rohe, W. Metzner, Eur. Phys. J. B. {\bf 48}, 349 (2005); R. Gersch, J. Reiss, C. Honerkamp, New J. Phys. \textbf{8}, 320 (2006); R. Gersch, C. Honerkamp, and W. Metzner, New J. Phys. \textbf{10}, 045003 (2008).

\bibitem{2009:Husemann} C. Husemann and M. Salmhofer, Phys. Rev.~B~\textbf{79}, 195125~(2009).

\bibitem{Katanin} A. Katanin, Phys. Rev. B \textbf{81}, 165118 (2010).

\bibitem{2011:Igoshev} P. A. Igoshev, V. Yu. Irkhin, and A. A. Katanin, Phys. Rev. B \textbf{83}, 245118 (2011).

\bibitem{Yuki} A. A. Katanin, H. Yamase, V. Yu. Irkhin, J. Phys. Soc. Jpn. {\bf 80}, 063702 (2011).

\bibitem{2013:Eberlein} A.~Eberlein and W.~Metzner, Phys. Rev.~B~\textbf{87}, 174523~(2013).

\bibitem{2017:TUfRG} J.~Lichtenstein, D. S\'anchez de~la~Pena, D.~Rohe, E.~Di~Napoli, C.~Honerkamp, S.~A. Maier, Computer Physics Communications~\textbf{213}, 100~(2017)

\bibitem{2022:BFfRG}
P.~M.~Bonetti, A.~Toschi, C.~Hille, S.~Andergassen, and D.~Vilardi, Phys. Rev. Research \textbf{4}, 013034~(2022).




\bibitem{2007:Igoshev} P. A. Igoshev, A. A. Katanin, and V. Yu. Irkhin, Journ. Exp. Theor. Phys. \textbf{105}, 1043 (2007).

\bibitem{2009:Igoshev} P. A. Igoshev, A. A. Katanin, H. Yamase, V. Yu. Irkhin, J. Magn. Magn. Mat. \textbf{321}, 899 (2009).

\bibitem{1994:Andersen}
P.~E.~Bl\"ochl, O.~Jepsen, O.~K.~Andersen, Phys.~Rev.~B~\textbf{49}, 16223 (1994).

\bibitem{1993:Berntsen}
J.~Berntsen, R.~Cools, and T.~O.~Espelid, ACM Transactions on Mathematical Software \textbf{19}, 320 (1993).

\bibitem{OurRev} G. Rohringer, H. Hafermann, A. Toschi, A. A. Katanin, A. E. Antipov, M. I. Katsnelson, A. I. Lichtenstein, A. N. Rubtsov, K. Held, Rev. Mod. Phys. {\bf 90}, 025003 (2018).

\bibitem{DMF2RG} C. Taranto, S. Andergassen, J. Bauer, K. Held, A. Katanin, W. Metzner, G. Rohringer, and A. Toschi, Phys. Rev. Lett. {\bf 112}, 196402 (2014); N. Wentzell, C. Taranto, A. A. Katanin, A. Toschi, and S. Andergassen, Phys. Rev. B {\bf 91}, 045120 (2015).






\end{thebibliography}
\end{document}